\documentclass[twocolumn,secnumarabic,amssymb, nobibnotes, superscriptaddress, nofootinbib,aps,prd]{revtex4-2}
\usepackage[colorlinks,citecolor=blue,linkcolor=blue,anchorcolor=blue,filecolor=blue, urlcolor=blue]{hyperref}
\usepackage{natbib}

\usepackage{amsmath}
\usepackage{adjustbox}
\usepackage{graphicx}
\usepackage{booktabs}       
\usepackage{amsfonts}       
\usepackage{nicefrac}       
\usepackage{microtype}
\usepackage{cleveref}
\usepackage{xcolor}
\usepackage{multirow}
\usepackage{makecell}
\usepackage{array}
\usepackage{orcidlink}
\usepackage{float}

\begin{document}

\title{Dark matter heating in evolving protoneutron stars: A Two-Fluid Approach}

\author{Adamu Issifu \orcidlink{0000-0002-2843-835X}}
\email{ai@academico.ufpb.br}
\affiliation{Departamento de F\'isica, Instituto Tecnol\'ogico de Aeron\'autica, DCTA, 12228-900, S\~ao Jos\'e dos Campos, SP, Brazil}
\affiliation{Laborat\'orio de Computa\c c\~ao Cient\'ifica Avan\c cada e Modelamento (Lab-CCAM), S\~ao Jos\'e dos Campos, SP, Brazil}
\affiliation{CFisUC, Department of Physics, University of Coimbra, 3004-516 Coimbra, Portugal}

\author{Prashant Thakur \orcidlink{0000-0003-4189-6176}}
\email{prashantthakur1921@gmail.com}
\affiliation{Department of Physics, Yonsei University, Seoul, 03722, South Korea}

\author{Davood~Rafiei Karkevandi \orcidlink{0000-0002-6031-041X}} 
\email{davood.rafiei64@gmail.com}
\affiliation{Institute for Theoretical Physics, University of Wroc{\l}aw, Plac Maksa Borna 9, 50-204 Wroc{\l}aw, Poland}

\author{Franciele M. da Silva \orcidlink{0000-0003-2568-2901}} 
\email{franciele.m.s@ufsc.br}
\affiliation{Departamento de F\'isica, CFM - Universidade Federal de Santa Catarina, \\ Caixa Postal 5064, CEP 880.35-972, Florian\'opolis, SC, Brazil.}
\affiliation{Theoretical Astrophysics, Institute for Astronomy and Astrophysics, University of T\"{u}bingen, 72076 T\"{u}bingen, Germany}

\author{D\'ebora P. Menezes \orcidlink{0000-0003-0730-6689}}
\email{debora.p.m@ufsc.br}
\affiliation{Departamento de F\'isica, CFM - Universidade Federal de Santa Catarina, \\ Caixa Postal 5064, CEP 880.35-972, Florian\'opolis, SC, Brazil.}

\author{Y. Lim~\orcidlink{0000-0001-9252-823X}} 
\email{ylim@yonsei.ac.kr}
\affiliation{Department of Physics, Yonsei University, Seoul, 03722, South Korea}

\author{Tobias Frederico \orcidlink{0000-0002-5497-5490}} 
\email{tobias@ita.br}
\affiliation{Departamento de F\'isica, Instituto Tecnol\'ogico de Aeron\'autica, DCTA, 12228-900, S\~ao Jos\'e dos Campos, SP, Brazil}
\affiliation{Laborat\'orio de Computa\c c\~ao Cient\'ifica Avan\c cada e Modelamento (Lab-CCAM), S\~ao Jos\'e dos Campos, SP, Brazil}

\begin{abstract}
Neutron stars (NSs) provide a unique laboratory to probe dark matter (DM) through its gravitational imprint on stellar evolution. We use a two-fluid framework with non-annihilating, asymmetric DM, both fermionic and bosonic, that interacts with ordinary matter (OM) solely through gravity. Within this framework, we track protoneutron stars (PNSs) across their thermal and compositional evolution via quasi-static modeling over the Kelvin--Helmholtz cooling timescale. We uncover a distinct thermal signature: DM cores deepen the gravitational potential, compressing and heating the baryonic matter, while extended DM halos provide external support, leading to cooling of the stellar matter. In contrast, hyperons and other exotic baryons soften the equation of state similarly to DM cores but reduce, rather than increase, the temperature. DM thus alters both temperature and particle distribution profiles in ways that provide a clear diagnostic of its presence. DM cores also enhance compactness and shift hyperon onset, with the strongest effects during deleptonization and neutrino-transparent phases due to reduced neutrino pressure contributions. Consequently, this early thermal evolution, observable through supernova neutrino light curves and young pulsar cooling curves, offers a direct, testable probe of DM in NSs.
\end{abstract}

\maketitle

\section{Introduction}

Observational evidence indicates that DM plays a crucial role in shaping galaxy rotation curves \cite{1980ApJ...238..471R, 1981AJ.....86.1825B}, the formation of large-scale cosmological structures \cite{Springel:2005nw, Planck:2015fie}, and gravitational lensing phenomena \cite{Clowe:2006eq, Massey:2015dkw}. Despite decades of intensive experimental efforts, however, its direct detection through ground-based searches remains elusive~\cite{XENON:2018voc, Schumann:2019eaa}. This persistent challenge motivates the development of theoretical models that aim to predict how DM may interact with OM, as well as to constrain its possible mass and coupling strengths \cite{Bertone:2016nfn, Feng:2010gw}. Given the extreme densities reached in the cores of NSs and the unique advantage that astrophysical observations provide over terrestrial experiments, indirect searches for DM through NS modeling have become increasingly important~\cite{Bertone:2007ae, Kouvaris:2010jy, DelPopolo:2020hel, Bramante:2023djs}. 

NSs are particularly attractive laboratories for DM physics because, when born in regions of high DM density such as the Galactic Center or dwarf galaxies, they can efficiently capture and accumulate DM \cite{Garani:2018kkd, Gnedin:2003rj}. If sufficient amounts are trapped, the presence of DM can alter the global structure \cite{Das:2020vng, Shakeri:2022dwg, Konstantinou:2024ynd} of the star and its multimessenger signatures, including thermal evolution \cite{Bell:2018pkk, Issifu:2024htq, Busoni:2021zoe, Issifu:2025qqw, Issifu:2025gsq, Avila:2023rzj}, tidal deformability \cite{Karkevandi:2021ygv, Diedrichs:2023trk, Rutherford:2024uix, RafieiKarkevandi:2021hcc, Collier:2022cpr,Sen:2022pfr, Guha:2024pnn}, and gravitational-wave emission~\cite{Giangrandi:2025rko}. The capture process occurs as DM particles scatter off baryons in the dense stellar medium. Since baryonic matter dominates the NS core, incoming DM particles experience successive collisions that dissipate their kinetic energy, eventually leading to gravitational capture and confinement inside the star. Over time, this accumulation may significantly affect the internal composition, stability, and observable properties of NSs. Consequently, modeling DM–NS matter interactions provides a promising pathway to constrain the DM mass, scattering cross section, and possible couplings with OM \cite{McKeen:2023ztq}. 

The precise nature of DM and its couplings to OM remain uncertain due to our limited knowledge of its fundamental properties \cite{Grippa:2024ach}. In the context of NSs, two main frameworks have been developed to model the possible presence of DM. In the two-fluid approach, DM is assumed to interact with OM solely through gravity, leading to a coexistence of two distinct fluids within the star~\cite{Kain:2021hpk, Thakur:2023aqm, RafieiKarkevandi:2021hcc, Karkevandi:2024vov,Pitz:2024xvh}. In contrast, the single-fluid approach assumes that DM interacts with OM via Standard Model mediators, such as the Higgs or other portal interactions, thereby modifying the effective EOS of the stellar matter~\cite{Lopes:2024ixl, Issifu:2025qqw, Panotopoulos:2017idn, Hajkarim:2024ecp, Shirke:2024ymc, Kumar:2025ytm, Lenzi:2022ypb,Shahrbaf:2022upc,Shahrbaf:2024gdm,Thakur:2024btu,Sen:2024yim}. How DM accumulates in NSs and its resulting impact on stellar properties depends strongly on the nature of the DM particle, whether fermionic or bosonic, and its interaction with OM. For instance, self-annihilating DM can be captured and subsequently annihilated inside the star, releasing energy that induces internal heating \cite{Jungman:1995df, Fermi-LAT:2015att, Kouvaris:2010vv}. This process alters the thermal evolution of NSs, potentially affecting the determination of their ages from standard cooling curves. On the other hand, non-self-annihilating DM candidates, such as asymmetric DM \cite{Petraki:2013wwa, Zurek:2013wia}, can accumulate over time without annihilation, leading to significant modifications of the global structure of the star, its maximum mass, oscillation spectra, tidal deformability, and gravitational-wave signatures \cite{Leung:2022wcf}. Together, these scenarios illustrate how NSs can serve as natural laboratories to probe the properties of DM across both gravitational and particle physics frontiers.

This work examines the impact of {asymmetric} DM on the thermal evolution of PNSs, assuming that DM is already present at the onset of PNS formation and persists through to the stage of cold, catalyzed NS configuration. This assumption is motivated by the expectation that progenitors located in high-density DM environments, such as the Galactic Center or dense dwarf galaxies, can accumulate appreciable amounts of DM before collapse~\cite{Gnedin:2003rj}. Moreover, since the characteristic timescale for DM capture and accumulation ($10^6$--$10^9$ years) \cite{McKeen:2023ztq, Grippa:2024ach} is many orders of magnitude longer than the short dynamical timescale of PNS evolution (tens of seconds) \cite{Prakash:1996xs, Janka:2006fh, Janka:2012wk, Pons:1998mm}, the DM content can be regarded as effectively fixed during this stage. Within this framework, we model the quasi-static evolution of PNSs using the two-fluid formalism, which enables a consistent analysis of the gravitational and thermodynamic effects introduced by the presence of DM.  We note, however, that the actual DM content is strongly dependent on the progenitor's environment and the assumed interaction cross section, and thus remains a model-dependent parameter \cite{Guver:2012ba}.

The DM model considered here incorporates both fermionic DM, derived from the mirrored model in \cite{Issifu:2024htq}, and self-interacting bosonic DM as described in \cite{Shakeri:2022dwg, Karkevandi:2021ygv}. The OM EOS is constructed within the relativistic mean field (RMF) framework, calibrated with the DDME2 parameterization to ensure agreement with astrophysical constraints and nuclear saturation properties \cite{PhysRevC.71.024312}. Applications of the RMF model with the DDME2 parameterization to PNSs are reported in \cite{Issifu:2023qyi, Malfatti:2019tpg, daSilva:2025cfe, Issifu:2025gsq}. This work presents a systematic study of the impact of DM on evolving NSs, addressing both its role in the stellar core and the formation of a DM halo, and analyzing their consequences for the thermal evolution of the star. These effects are modeled using the two-fluid Tolman–Oppenheimer–Volkoff (TOV) approach with non-interacting fluids, assuming a fixed DM mass fraction from the star's birth to its maturity. In \cite{Issifu:2024htq}, DM-admixed PNSs were studied under the assumption that OM and mirror DM are in thermal equilibrium. Here, we investigate whether the heating observed, previously attributed to the presence of DM, originates from the thermal equilibrium assumption, from the gravitational influence of DM, or from the specific properties of the DM component. Our working assumption is that the star is born in a dense DM environment but evolves without reaching thermal equilibrium with the DM; that is, DM is assumed to be at $T=0$ MeV in all evolutionary stages. In our model, the two fluids interact solely through gravity, which ensures a coupling between them while ruling out any direct non-gravitational thermalization \cite{Bertone:2004pz}.

The work unveils a unique, gravitationally driven thermal signature of DM in NSs. We demonstrate that non-annihilating DM can heat the stellar core, a result that directly contrasts with the cooling induced by exotic baryons like hyperons. This heating arises from gravitational compression: a DM core deepens the potential well, raising the baryonic temperature to maintain equilibrium, while an extended halo provides external support, cooling the star. Furthermore, we present the first systematic study of Bose-Einstein condensate DM within PNSs using a two-fluid approach. This thermal fingerprint provides a clear, testable discriminant for DM, with direct implications for interpreting supernova neutrino light curves and the cooling tracks of young pulsars.

We choose three representative DM mass fractions for our analysis. However, we stress that the DM mass fractions adopted here (1\%, 5\%, and 10\%) are not intended to represent realistic accumulation levels through gravitational capture, which are estimated to be far below $10^{-10}\,M_{\odot}$ \cite{Bramante:2023djs}. Rather, these values serve as a hypothetical exploration of how DM could influence the structure and evolution of PNSs, following conventions commonly adopted in the literature \cite{Cipriani:2025tga, Emma:2022xjs, Konstantinou:2024ynd}. Alternative mechanisms, such as neutron dark decays \cite{PhysRevLett.120.191801}, could in principle generate larger internal DM populations, although such scenarios lie beyond the purely gravitational framework adopted here.

The paper is structured as follows. In \cref{fm}, we present the theoretical framework, beginning with \cref{bm}, which details the EOS of the OM sector. \Cref{dm} introduces the dark sector models, with \cref{fd} focusing on fermionic DM and \cref{bd} on self-interacting bosonic DM. This section concludes with the two-fluid TOV formalism in \cref{ttov}. The results and analysis are presented in \cref{ra}, covering the microscopic properties of stellar matter \cref{msp}, the mass–radius relation \cref{mr}, the temperature profile \cref{tp}, and the features arising solely from DM \cref{dms}. Finally, \cref{fr} provides our concluding remarks.

\section{Model Formalism}\label{fm}
The study considers three distinct models: the fermionic mirror DM model, the self-interacting bosonic DM model, and the RMF model for the visible sector. The details of each are discussed in the following subsections.

\subsection{Baryonic Matter}\label{bm}

\begin{table}[h]
\begin{center}
\caption {{DDME2 parameters for nucleonic and hyperonic matter used in this work. The model parameters are $a_i$, $b_i$, $c_i$, and $d_i$, and $g_{iN}(n_0)$ denote the meson--nucleon (hyperon) couplings at saturation density \cite{PhysRevC.71.024312, Typel:2018cap}.}}
\begin{tabular}{ c c c c c c c }
\hline
 meson($i$) & $m_i(\text{MeV})$ & $a_i$ & $b_i$ & $c_i$ & $d_i$ & $g_{i N} (n_0)$\\
 \hline
 \hline
 $\sigma$ & 550 & 1.3881 & 1.0943 & 1.7057 & 0.4421 & 10.5396 \\  
 $\omega$ & 783 & 1.3892 & 0.9240 & 1.4620 & 0.4775 & 13.0189  \\
 $\rho$ & 763 & 0.5647 & --- & --- & --- & 7.3672 \\
 $\phi_0$ &1019&1.3892 & 0.9240 & 1.4620 & 0.4775 & 13.0189\\
 \hline
 \hline
\end{tabular}
\label{T1}
\end{center}
\end{table}

\begin{table}[h]
\begin{center}
\caption {The ratio of the baryon coupling to the corresponding nucleon coupling for hyperons. 
}
\label{T2}
\begin{tabular}{ c c c c c } 
\hline
\hline
 b & $\chi_{\omega b}$ & $\chi_{\sigma b}$ & $\chi_{\rho b}$ & $\chi_{\phi b}$  \\
 \hline
 $\Lambda$ & 0.714 & 0.646 & 0 & -0.808  \\  
$\Sigma^0$ & 1 & 0.663 & 0 & -0.404  \\
  $\Sigma^{-}$, $\Sigma^{+}$ & 1 & 0.663 & 1 & -0.404  \\
$\Xi^-$, $\Xi^0$  & 0.571 & 0.453 & 0 & -1.01 \\
  \hline
  \hline
\end{tabular}
\label{T2}
\end{center}
\end{table}
The OM sector is modeled using the RMF approach, where the dynamics of baryons, mesons, and leptons are governed by the Lagrangian density
\begin{equation}\label{rmf}
     \mathcal{L}_{\rm RMF}= \mathcal{L}_{H}+ \mathcal{L}_{\rm m}+ \mathcal{L}_{\rm l},
\end{equation}
with $\mathcal{L}_H$, $\mathcal{L}_m$, and $\mathcal{L}_{\rm l}$ representing the Lagrangian densities of the baryon octet, the meson fields, and the leptons, respectively. Explicitly, the baryonic Lagrangian reads
\begin{align}
 \mathcal{L}_{H}= {}& \sum_{b\in H}  \bar \psi_b \Big[  i \gamma^\mu\partial_\mu - \gamma^0  \big(g_{\omega b} \omega_0  +  g_{\phi b} \phi_0+ g_{\rho b} I_{3b} \rho_{03}  \big)\nonumber \\
 &- \Big( m_b- g_{\sigma b} \sigma_0 \Big)  \Big] \psi_b,
\end{align}
where $\psi_b$ and $m_b$ denote the baryon field and mass, $g_{i b}$ are the coupling constants to the meson fields $i=\sigma, \omega, \phi, \rho$, and $I_{3b}$ is the isospin projection of the baryon.  The meson Lagrangian is given by
\begin{equation}\label{lagrangian}
 \mathcal{L}_{\rm m} = - \frac{1}{2} m_\sigma^2 \sigma_0^2  + \frac{1}{2} m_\omega^2 \omega_0^2 + \frac{1}{2} m_\phi^2 \phi_0^2 + \frac{1}{2} m_\rho^2 \rho_{03}^2,
\end{equation}
describing the mean-field contributions of the scalar and vector mesons to the system.  

The lepton contribution is
\begin{equation}\label{l1}
 \mathcal{L}_{\rm l} = \sum_l \bar \psi_l \left( i\gamma^\mu \partial_\mu - m_l \right) \psi_l,
\end{equation}
where $\psi_l$ and $m_l$ denote the lepton fields and masses, included to ensure charge neutrality and beta-equilibrium in the stellar matter.  This framework allows for the calculation of the baryonic energy density, pressure, and composition under the RMF approach, which are then used in the two-fluid TOV equations when including the DM sector. The EOS is computed at fixed entropy and lepton fractions corresponding to different stages of stellar evolution on the Kelvin–Helmholtz timescale, allowing us to model the thermal and compositional evolution of PNSs. A full derivation of the energy density and pressure at finite temperature, applicable to PNS studies, can be found in \cite{Issifu:2023qyi, Issifu:2024htq, Issifu:2024fuw}. The coupling constants for the DDME2 parameters and for the hyperons are presented in \cref{T1} and \cref{T2}, respectively, with $\chi_{ib} = g_{ib}/g_{iN}$ denoting the ratio of the hyperon coupling to the corresponding nucleon coupling. The following subsections describe the DM models employed in this study.

\subsection{Dark Matter Models}\label{dm}

In this work, we consider two types of DM candidates: a self-interacting bosonic DM capable of forming a Bose–Einstein condensate (BEC) and a fermionic mirror DM capable of forming a dark NS in isolation. In the fermionic DM subsection below, we focus on core configurations that serve as a basis for comparison with the bosonic core configurations analyzed in this paper. This approach allows us to extend our previous work \cite{Issifu:2024htq}, which explored core configurations of fermionic mirror DM admixed with evolving PNSs. Additionally, we examine a bosonic DM halo scenario to assess how the extended DM distribution influences the cooling process of protoneutron stars.

\subsubsection{Fermionic Mirror Dark Matter}\label{fd}
The fermionic DM model is described by the Lagrangian density \cite{Xiang:2013xwa, Thakur:2023aqm}:
\begin{align}\label{L1}
    {\cal L}_{\rm DM} = {} & \overline\psi_D \Big[ i \gamma_\mu \partial^\mu - \gamma^0 g_v V_0 - (m_D - g_{\tilde{\sigma}} \tilde{\sigma}_0) \Big] \psi_D \nonumber\\
    & - \frac{1}{2} m_{\tilde{\sigma}}^2 \tilde{\sigma}_0^2 + \frac{1}{2} m_v^2 V_0^2,
\end{align}
where $\psi_D$ and $m_D$ denote the fermionic DM field and its mass. The mean-field vector and scalar mesons, $V_0$ and $\tilde{\sigma}_0$, mediate interactions between DM particles, with coupling constants $g_v$ and $g_{\tilde{\sigma}}$ and masses $m_v$ and $m_{\tilde{\sigma}}$, respectively \cite{Issifu:2024htq}. The coupling strengths and meson masses are the same as the OM sector in \cref{T1}.

The model permits a self-consistent determination of the DM energy density, pressure, and number density under the conditions of dark charge neutrality and dark $\beta$-equilibrium, the latter introduced by free dark-sector leptons in analogy with \cref{l1}. The formalism is discussed in detail in \cite{Issifu:2024htq}; here, we adopt the same framework but in $T=0$~MeV setup.

\subsubsection{Self-interacting Bosonic DM}\label{bd}
The bosonic DM model is described by a self-interacting scalar field with the Lagrangian density:
\begin{equation}
    \mathcal{L}_{\rm BDM} = \frac{1}{2} \partial_\mu \phi^* \partial^\mu \phi - \frac{1}{2} m_\phi^2 \phi^* \phi - \frac{\lambda}{4} (\phi^* \phi)^2,
\end{equation}
where $\phi$ is the complex scalar DM field, $m_\phi$ is the mass of the bosonic DM particle, and $\lambda$ is the self-interaction coupling constant. The quartic term $(\phi^* \phi)^2$ accounts for repulsive self-interactions among bosonic DM particles, which can support a higher stellar mass.
In the mean-field approximation, the scalar field is replaced with their ground-state expectation value, $\phi \rightarrow \langle \phi \rangle = \phi_0$. Moreover, at sufficiently low temperatures, the scalar field is expected to form a BEC \cite{Liebling:2012fv,Visinelli:2021uve}. In this work, we assume $T=0$~MeV, which eliminates thermal fluctuations and allows for the formation of a BEC. The corresponding EOS for the bosonic DM is given by \cite{Shakeri:2022dwg, Karkevandi:2021ygv}:
\begin{equation}\label{dmeq}
P_{\rm BDM} = \frac{m_\phi^4}{9 \lambda} \left( \sqrt{1 + \frac{3 \lambda \varepsilon_{\rm BDM}}{m_\phi^4}} - 1 \right)^2,
\end{equation}
where $P_{\rm BDM}$ and $\varepsilon_{\rm BDM}$ are the pressure and energy density of the bosonic DM fluid, respectively.

This formulation allows the bosonic DM to form either a dense core or an extended halo around the baryonic matter based on the choice of $m_\phi$ and $\lambda$ \cite{Shakeri:2022dwg}. The bosonic DM fluid is incorporated into the two-fluid TOV equations, interacting with the OM solely through gravity. This model provides a flexible framework to study the structural and thermal impact of bosonic DM on evolving PNSs and the final formation of cold, catalyzed NSs configuration, complementing the analysis performed with fermionic DM. Detailed derivations and the determination of the suitable coupling constants can be found in~\cite{Shakeri:2022dwg, Karkevandi:2021ygv}, and references therein. For this study, we choose $m_\phi = 500 \,\rm MeV$ for the core case and $100\,\rm MeV$ for the halo case, with a coupling constant $\lambda = \pi$. It is worth mentioning that, for sub-GeV bosons, obtaining stellar-mass configurations requires a self-coupling of order $\lambda \sim 0.1-10$, while perturbativity and unitarity demand $\lambda \lesssim 4\pi$. We therefore adopt $\lambda=\pi$ as a representative benchmark and vary only the boson mass, taking $m_\phi=100~\mathrm{MeV}$ for the DM halo case and $m_\phi=500~\mathrm{MeV}$ for DM core formation. It should also be noted that one could equivalently fix the mass and vary the coupling; however, the qualitative effect on the formation of DM halos versus DM cores remains the same. 

{ 
It should be mentioned that, for a fixed coupling constant $\lambda$, the DM particle mass $m_\phi$ determines the stiffness of the EOS. 
Small $m_\phi$ values correspond to a \emph{stiff} DM EOS, favoring extended halos, whereas large $m_\phi$ values lead to a \emph{soft} EOS that produces compact cores. 
In different density limits, one can approximate Eq.~(\ref{dmeq}) by a polytrope $P=K\varepsilon^\gamma$: at low density $\gamma\simeq2$, and it smoothly approaches $\gamma\simeq1$ at high density. 
At low density, Eq.~(\ref{dmeq}) reduces to
\begin{equation}
P_{\rm BDM} \approx \frac{\lambda}{4m_{\phi}^4}\,\varepsilon_{\rm BDM}^{2},
\end{equation}
where $K=\lambda/(4m_\phi^4)$. 
In the high-density (or equivalently, very light-mass or strong-coupling) regime, Eq.~(\ref{dmeq}) approaches the radiation-like form $P\simeq \epsilon/3$~\cite{RafieiKarkevandi:2021hcc}. The transition between these regimes occurs when $\lambda\,\varepsilon_{\rm BDM}\sim m_\phi^4$ (i.e.\ $3\lambda\varepsilon_{\rm BDM}/m_\phi^4\sim\mathcal{O}(1)$), signaling the transition from the quadratic regime to the radiation-like regime.
}

\subsection{Two-Fluid TOV Formalism}\label{ttov}

In this work, the interaction between the dark sector and the OM sector is assumed to be purely gravitational. Consequently, the coupling between the two fluids is introduced solely through the two-fluid TOV equations, expressed as \cite{Das:2020ecp, Rutherford:2022xeb, Shawqi:2024jmk, Biesdorf:2024dor, Karkevandi:2024vov}:  
\begin{align}
\frac{dP_{\rm OM}}{dr} ={}& -\left(P_{\rm OM} + \varepsilon_{\rm OM}\right)\frac{4\pi r^3\left(P_{\rm OM}+P_{\rm D}\right)+M(r)}{r\left(r-2M(r)\right)}, \label{tov}\\
\frac{dP_{\rm D}}{dr} ={}& -\left(P_{\rm D} + \varepsilon_{\rm D}\right)\frac{4\pi r^3\left(P_{\rm OM}+P_{\rm D}\right)+M(r)}{r\left(r-2M(r)\right)}, \label{tov1}
\end{align}
with the enclosed mass profile determined by
\begin{equation}\label{tov2}
\frac{dM(r)}{dr} = 4\pi \left(\varepsilon_{\rm OM} + \varepsilon_{\rm D}\right) r^2.
\end{equation}
Here, $P_{\rm OM}$ and $\varepsilon_{\rm OM}$ denote the pressure and energy density of the baryonic sector, while $P_{\rm D}$ and $\varepsilon_{\rm D}$ correspond to those of the dark sector. The total gravitational mass, $M(R) = M_{\rm OM}(R_{\rm OM}) + M_D(R_D)$, 
of the configuration naturally emerges from the combined contributions of both fluids.  

Since we assume a fixed fraction of DM to be present from the birth of the star through its evolutionary stages, we define the DM mass fraction as
\begin{equation}
F_{\rm D} = \frac{M_{\rm D}(R_{\rm D})}{M(R)},
\end{equation}
where $M_{\rm D}$ and $R_{\rm D}$ represent the gravitational mass and radius of the DM component, and $M$ is the total gravitational mass (baryonic + dark) of the star, and $R$ is the outer radius of the star. $M_{\rm OM}$ and $R_{\rm OM}$ are the mass and radius of the OM sector, respectively. For core configurations, $R$ denotes the OM sector, while for halo configurations, it refers to the dark sector. This parametrization offers a unified measure of DM content across stellar evolution, enabling direct comparison with purely baryonic stars. 

The conserved baryonic mass of the OM component in the two-fluid framework is given by
\begin{equation}\label{bm}
M_B = 4\pi M_N \int_0^{R_B} n_B(r) \left(1-\dfrac{2M(r)}{r} \right)^{-1/2} r^2dr,
\end{equation}
where $M_N = 938~\rm MeV$ is the nucleon rest mass, $n_B$ is the baryon density, $M$ is the total gravitational mass defined in \cref{tov2}, and $R_B$ is the radius of the baryonic fluid. The baryonic mass $M_B$ is a conserved quantity throughout the star's evolution, since it directly tracks the number of baryons, unaffected by thermal processes or gravitational binding. The difference between $M_B$ and the gravitational mass $M$ defines the total binding energy, $E_b = (M_B - M)$, which measures the energy released during stellar formation and plays a crucial role in determining the stability of compact configurations.  A larger binding energy generally indicates a more stable star, while the turning point of $M$ with respect to central density, combined with binding energy considerations, signals the onset of instability and possible collapse.

Although $F_D$ is kept fixed across evolutionary stages for consistency, the absolute DM mass varies because $F_D$ is defined relative to the total gravitational mass $M$, which evolves with the thermodynamic state of the star. Moreover, the conserved baryonic mass $M_B$ defined in \cref{bm} is not entirely independent of the DM component, since the metric factor entering the proper volume element depends on the total enclosed mass $M(r)$ contributed by both fluids. Thus, $M_B$ is gravitationally influenced by the DM sector \cite{Issifu:2025gsq} even though it tracks only the baryon number. A fully self-consistent treatment enforcing independent conservation of the DM particle number is deferred to future work.

\section{Results and Analysis}\label{ra}

\begin{table*}[t!]
\centering
\setlength\extrarowheight{2pt}
\begin{ruledtabular}
\caption{Properties of evolving PNSs with bosonic DM ($m_\phi=500~\rm MeV$) concentrated in their cores for nucleonic (N) configurations. The maximum gravitational mass ($M_{\rm max}$) and corresponding baryonic mass ($M_B$), radius ($R$), central energy density ($\varepsilon_{c}$), central baryon density ($n_c$), and core temperature ($T_c$) are shown in the table for a fixed lepton fraction $Y_l$ and entropy per baryon $s_B$.
}
\begin{tabular}{llcccccc}
\textbf{Model} & \textbf{$F_D$ [\%]} & \textbf{$M_B$}[$M_\odot$] & $M_{\rm max}$ [$M_\odot$] & $R$ [km] & $\varepsilon_{c}$ [MeVfm$^{-3}$] & $n_c$ [fm$^{-3}$] & \textbf{$T_c$} [MeV] \\
\hline
\multirow{4}{*}{$s_B=1$, $Y_l=0.4$ (N)}   
& No DM           & 2.80 & 2.44 & 12.20 & 1070 & 0.794 & 25.79 \\
& 1\% DM          & 2.75 & 2.40 & 12.17 & 1097 & 0.806 & 30.24 \\
& 5\% DM          & 2.55 & 2.28 & 11.78 & 1325 & 0.918 & 31.81 \\
& 10\% DM         & 2.35 & 2.11 & 11.50 & 1512 & 0.998 & 32.89 \\
\hline
\multirow{4}{*}{$s_B=2$, $Y_l=0.2$ (N)}   
& No DM           & 2.88 & 2.49 & 12.70 & 1007 & 0.759 & 60.76 \\
& 1\% DM          & 2.83 & 2.46 & 12.66 & 1035 & 0.773 & 72.36 \\
& 5\% DM          & 2.62 & 2.32 & 12.27 & 1247 & 0.876 & 76.23 \\
& 10\% DM         & 2.41 & 2.15 & 11.99 & 1422 & 0.954 & 78.89 \\
\hline
\multirow{4}{*}{$s_B=2$, $Y_{\nu_e}=0$ (N)}   
& No DM           & 2.90 & 2.49 & 12.75 & 1010 & 0.763 & 59.84 \\
& 1\% DM          & 2.85 & 2.46 & 12.69 & 1094 & 0.809 & 71.58 \\
& 5\% DM          & 2.64 & 2.33 & 12.29 & 1240 & 0.876 & 73.94 \\
& 10\% DM         & 2.43 & 2.16 & 12.01 & 1413 & 0.954 & 76.38 \\
\hline
\multirow{4}{*}{$T=0$ MeV (N)}           
& No DM           & 3.01 & 2.48 & 12.06 & 1050 & 0.813 & -- \\
& 1\% DM          & 2.95 & 2.45 & 11.96 & 1077 & 0.828 & -- \\
& 5\% DM          & 2.74 & 2.32 & 11.67 & 1305 & 0.939 & -- \\
& 10\% DM         & 2.51 & 2.15 & 11.28 & 1492 & 1.024 & -- \\
\end{tabular}
\label{tab:Mmax_R14_N}
\end{ruledtabular}
\end{table*}

\begin{table*}[t!]
\centering
\setlength\extrarowheight{2pt}
\begin{ruledtabular}
\caption{Properties of evolving PNSs with bosonic DM ($m_\phi=500~\rm MeV$) concentrated in their cores for hyperonic (H) configurations. The maximum gravitational mass ($M_{\rm max}$) and corresponding baryonic mass ($M_B$), radius ($R$), central energy density ($\varepsilon_{c}$), central baryon density ($n_c$), and core temperature ($T_c$) are shown in the table.}
\begin{tabular}{llcccccc}
\textbf{Model} & \textbf{$F_D$ [\%]} & \textbf{$M_B$}[$M_\odot$] & $M_{\rm max}$ [$M_\odot$] & $R$ [km] & $\varepsilon_{c}$ [MeVfm$^{-3}$] & $n_c$ [fm$^{-3}$] & \textbf{$T_c$} [MeV] \\
\hline
\multirow{4}{*}{$s_B=1$, $Y_l=0.4$ (H)}   
& No DM           & 2.65 & 2.33 & 12.20 & 1082 & 0.820 & 21.35 \\
& 1\% DM          & 2.60 & 2.30 & 12.05 & 1109 & 0.834 & 29.54 \\
& 5\% DM          & 2.41 & 2.17 & 11.78 & 1330 & 0.945 & 30.36 \\
& 10\% DM         & 2.21 & 2.01 & 11.47 & 1513 & 1.034 & 32.05 \\
\hline
\multirow{4}{*}{$s_B=2$, $Y_l=0.2$ (H)}   
& No DM           & 2.64 & 2.31 & 12.38 & 1080 & 0.823 & 46.48 \\
& 1\% DM          & 2.59 & 2.28 & 12.24 & 1112 & 0.844 & 59.17 \\
& 5\% DM          & 2.40 & 2.15 & 11.98 & 1334 & 0.956 & 62.08 \\
& 10\% DM         & 2.21 & 1.99 & 11.68 & 1513 & 1.043 & 64.10 \\
\hline
\multirow{4}{*}{$s_B=2$, $Y_{\nu_e}=0$ (H)}   
& No DM           & 2.62 & 2.28 & 12.30 & 1098 & 0.843 & 48.74 \\
& 1\% DM          & 2.58 & 2.25 & 12.21 & 1185 & 0.890 & 58.37 \\
& 5\% DM          & 2.41 & 2.13 & 11.94 & 1335 & 0.969 & 60.37 \\
& 10\% DM         & 2.20 & 1.97 & 11.54 & 1515 & 1.058 & 62.22 \\
\hline
\multirow{4}{*}{$T=0$ MeV (H)}           
& No DM           & 2.67 & 2.26 & 11.95 & 1085 & 0.864 & -- \\
& 1\% DM          & 2.63 & 2.23 & 11.82 & 1115 & 0.886 & -- \\
& 5\% DM          & 2.45 & 2.10 & 11.52 & 1340 & 1.009 & -- \\
& 10\% DM         & 2.23 & 1.94 & 11.10 & 1517 & 1.098 & -- \\
\end{tabular}
\label{tab:Mmax_R14_H}
\end{ruledtabular}
\end{table*}

\begin{table*}[t]
\centering
\renewcommand{\arraystretch}{1.2}
\caption{Properties of evolving PNSs with a dark matter halo are shown for a bosonic DM candidate of mass $m_\phi = 100~\rm MeV$, considering different DM fractions. Reported are the maximum gravitational mass ($M_{\max}$) and corresponding baryonic mass ($M_B$), dark matter halo radius ($R_D$), central energy density ($\varepsilon_c$), central baryon density ($n_c$), and core temperature ($T_c$).
}
\begin{tabular}{c c c c c c c c}
\hline
\textbf{Model} & \textbf{ $F_D$[\%]} & $M_B\,[M_\odot]$ & $M_{\max}\,[M_\odot]$ & $R_D$ [km] & $\varepsilon_c$ [MeV fm$^{-3}$] & $n_c$ [fm$^{-3}$] & $T_c$ [MeV] \\
\hline
\multirow{4}{*}{$s_B = 1,\; Y_{l} = 0.4$ (N)} 
 & No DM   & 2.80 & 2.44 & 12.20 & 1070 & 0.794 & 25.79 \\
 & 1\% DM  & 2.77 & 2.42 & 12.10 & 1088 & 0.802 & 30.24 \\
 & 5\% DM  & 2.70 & 2.37 & 19.40 & 1150 & 0.832 & 29.22 \\
 & 10\% DM & 2.75 & 2.40 & 34.24 & 1100 & 0.808 & 28.24 \\
\hline
\multirow{4}{*}{$s_B=2,\; Y_l=0.2$ (N)} 
 & No DM   & 2.88 & 2.49 & 12.70 & 1007 & 0.759 & 60.76 \\
 & 1\% DM  & 2.85 & 2.47 & 12.71 & 1026 & 0.769 & 71.09 \\
 & 5\% DM  & 2.78 & 2.43 & 19.82 & 1083 & 0.796 & 69.85 \\
 & 10\% DM & 2.83 & 2.46 & 34.48 & 1035 & 0.773 & 67.39 \\
\hline
\multirow{4}{*}{$s_B = 2,\; Y_{\nu_e} = 0.0$ (N)} 
 & No DM   & 2.90 & 2.49 & 12.75 & 1010 & 0.763 & 59.84 \\
 & 1\% DM  & 2.86 & 2.47 & 13.50 & 1071 & 0.796 & 70.42 \\
 & 5\% DM  & 2.80 & 2.43 & 20.61 & 1129 & 0.825 & 68.14 \\
 & 10\% DM & 2.85 & 2.46 & 35.28 & 1102 & 0.812 & 65.91 \\
\hline
\multirow{4}{*}{$T = 0$ MeV (N)} 
 & No DM   & 3.01 & 2.48 & 12.06 & 1050 & 0.813 & -- \\
 & 1\% DM  & 2.97 & 2.46 & 12.00 & 1069 & 0.824 & -- \\
 & 5\% DM  & 2.89 & 2.42 & 18.47 & 1140 & 0.859 & -- \\
 & 10\% DM & 2.92 & 2.43 & 33.59 & 1107 & 0.843 & -- \\
\hline
\hline
\end{tabular}
\label{tab:nucleons_filled}
\end{table*}

This section is divided into three subsections to examine the net impact of DM on both the microscopic and macroscopic properties of NSs, as well as on the thermal evolution of young stars from birth through the supernova phase. The main results are summarized in \cref{tab:Mmax_R14_N} and \cref{tab:Mmax_R14_H} for the core-DM configuration and in \cref{tab:nucleons_filled} for the halo configuration, providing key values for convenient comparison. \Cref{tab:Mmax_R14_N} shows a systematic decrease in the $M_{\rm max}$ and $R$ as the $F_D$ increases across evolutionary stages. This trend arises because a higher $F_D$ intensifies the gravitational field, compresses the baryonic matter, and redistributes mass, leading to smaller $M_B$, $M_{\rm max}$, and stellar $R$. The resulting compactification raises the $n_c$ and $\varepsilon_c$, which in turn increases the $T_c$, as reflected in the table. The inclusion of hyperons, \cref{tab:Mmax_R14_H}, also leads to a reduction in $M_B$, $M_{\rm max}$, and $R$. However, it does not increase $T_c$, instead, the central temperature decreases because the thermal energy of the PNS is conserved. Introducing an additional baryonic degree of freedom spreads the thermal energy over more particles, reducing the thermal energy per particle.

\Cref{tab:nucleons_filled} shows the halo configuration with purely nucleonic matter composition for comparison. Relative to the no-DM baseline, both $M_{\rm max}$ and $M_B$ increase once $F_D > 5\%$, indicating that a sufficiently massive halo can add gravitational binding without significantly disrupting the nucleonic core. The central energy density $\varepsilon_c$ exhibits the opposite trend: it rises for small halos, where the DM slightly compresses the baryonic core, but decreases for larger $F_D$, where the extended halo dilutes the average density of the configuration. The halo systematically increases the stellar radius, as the presence of DM outside the baryonic core shifts mass distribution outward and provides additional pressure support in the outer layers, except for the 1\% case, in the first and the last stages, where the star is more compact. The central density $n_c$ grows with increasing $F_D$, reflecting compression of the inner core by the gravitational pull of the halo, but it drops slightly at $F_D = 10\%$, where the DM mass dominates the outer layers and reduces central compactification. The thermal behavior shows that while DM initially raises $T_c$, a distinctive feature of DM admixture caused by gravitational compression of the core, for $F_D > 1\%$ the halo begins to draw thermal energy outward. This results in a systematic reduction of $T_c$, consistent with the redistribution of entropy and pressure support from the baryonic core to the DM halo. Overall, this behavior reflects how halo DM not only modifies the global structure (mass–radius relation) but also alters the star's thermal balance by effectively stiffening the EOS and redistributing energy. Generally, it is worth mentioning that in two-fluid DM admixed stars, DM can mimic an effective softening or stiffening of the global EOS, as reflected in changes of the mass–radius relation. Indeed, a DM core typically reduces M and R
(softening-like), while an extended DM halo can increase the total radius and mass (stiffening-like), thus modifying the global EOS.

\subsection{Microscopic Properties of the Stellar Matter}\label{msp}

\begin{figure*}
    \centering
    \includegraphics[width=1\linewidth]{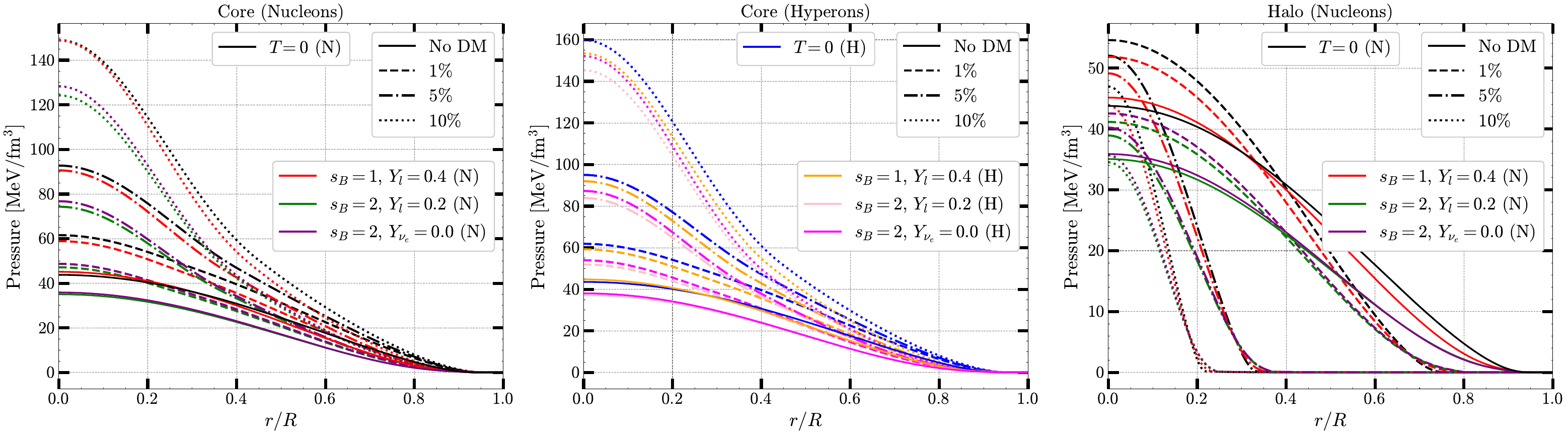}
    \caption{Variation of the core pressure of a star with a fixed baryon mass of $1.55\,M_\odot$ as a function of the normalized radius $r/R$ during the evolution of PNSs admixed with bosonic DM. The left and middle panels show stars with a fixed DM fraction in the core ($m_\phi = 500$~MeV), while the right panel shows stars with a DM halo ($m_\phi = 100$~MeV) at various stages of stellar evolution. The color codes represent different thermodynamic stages, while the line styles indicate the DM fractions.
    }
    \label{fig:Pressure_radius_core_halo}
\end{figure*}

Determining the microscopic properties of DM-admixed NS matter within the two-fluid formalism is challenging, as DM and OM interact solely through gravity. While each component obeys its own EOS, the total pressure and energy density are coupled through spacetime curvature. We therefore extract a configuration-dependent effective pressure profiles from the two-fluid TOV solutions. Although not a fundamental thermodynamic EOS, since it depends on the $F_D$ and global structure, it provides a useful diagnostic of how gravitational coupling modifies pressure support in DM-admixed stars.
In \cref{fig:Pressure_radius_core_halo}, we show these profiles for a star with fixed baryonic mass $1.55\,\rm M_\odot$ across all evolutionary stages, from the hot, lepton-rich PNS to the final cold, catalyzed NS. This framework allows us to track how thermodynamic changes, composition, and DM content reshape the core pressure throughout stellar evolution.

The three panels consider a bosonic DM candidate capable of forming a Bose–Einstein condensate \cite{Pethick:2015jma, Shakeri:2022dwg, Karkevandi:2021ygv} under both core and halo-like configurations \cite{Cermeno:2016olb}. The first panel depicts nucleonic stars with a compact DM core, the second includes hyperonic matter with DM core, and the third illustrates an extended DM halo surrounding the baryonic star. Comparing the first two panels, the inclusion of hyperons slightly increases the core pressure due to the additional degrees of freedom, which softens the EOS and enhances stellar compactness \cite{Balberg:1997yw}.

Examining the first two panels in detail, the core pressure peaks during the neutrino-trapping stage when the star is lepton-rich. After core bounce, it drops to a minimum during deleptonization, driven by shock propagation and neutrino diffusion that increase the $s_B$, reduce the $Y_l$, and raise the thermal energy. This expands the stellar radius and softens the EOS, lowering the central pressure. As neutrinos escape, the star gradually contracts, producing a modest rise in pressure, which becomes more pronounced as the star cools into the final cold, catalyzed NS, surpassing values obtained during the core birth. For core-admixed DM configurations, the gravitational contribution of DM increases stellar compactness, raising the core pressure relative to the no-DM baseline, with the increase following the order 1\%, 5\%, and 10\% of $F_D$.

In contrast, the extended DM halo in the third panel redistributes gravity outward, reducing the pressure requirement on the baryonic core. Consequently, the core pressure is significantly lower than in the core-DM case for the same stellar composition (compare first and third panels). Interestingly, the core pressure decreases with increasing halo fraction, being the highest for a 1\% halo, lower for a 5\% halo, and the lowest for a 10\% halo, with all being above the no-DM baseline. As the halo grows, it provides more gravitational support to the star, relieving the baryonic core and reducing its central pressure. Together, these results illustrate the contrasting effects of DM distribution: core-admixed DM deepens the central potential and heats the star (see Figs.~\ref{fig:temp_radius_core} and \ref{Temperature_radius_Halo} below), while a diffuse halo shifts gravitational support outward, lowering core pressure.

\begin{figure*}[!t]
\centering
\includegraphics[width=0.49\linewidth]{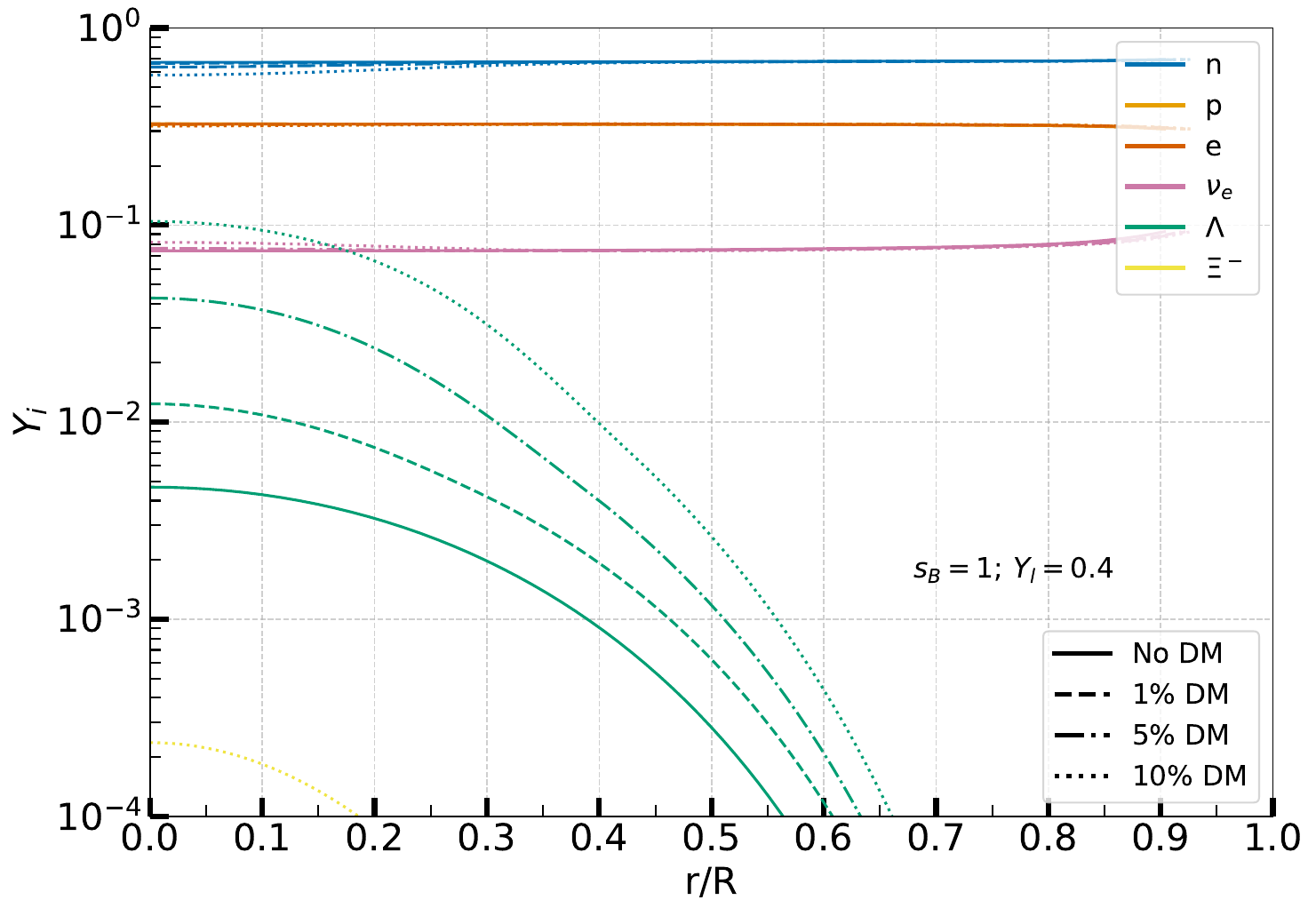}
\includegraphics[width=0.49\linewidth]{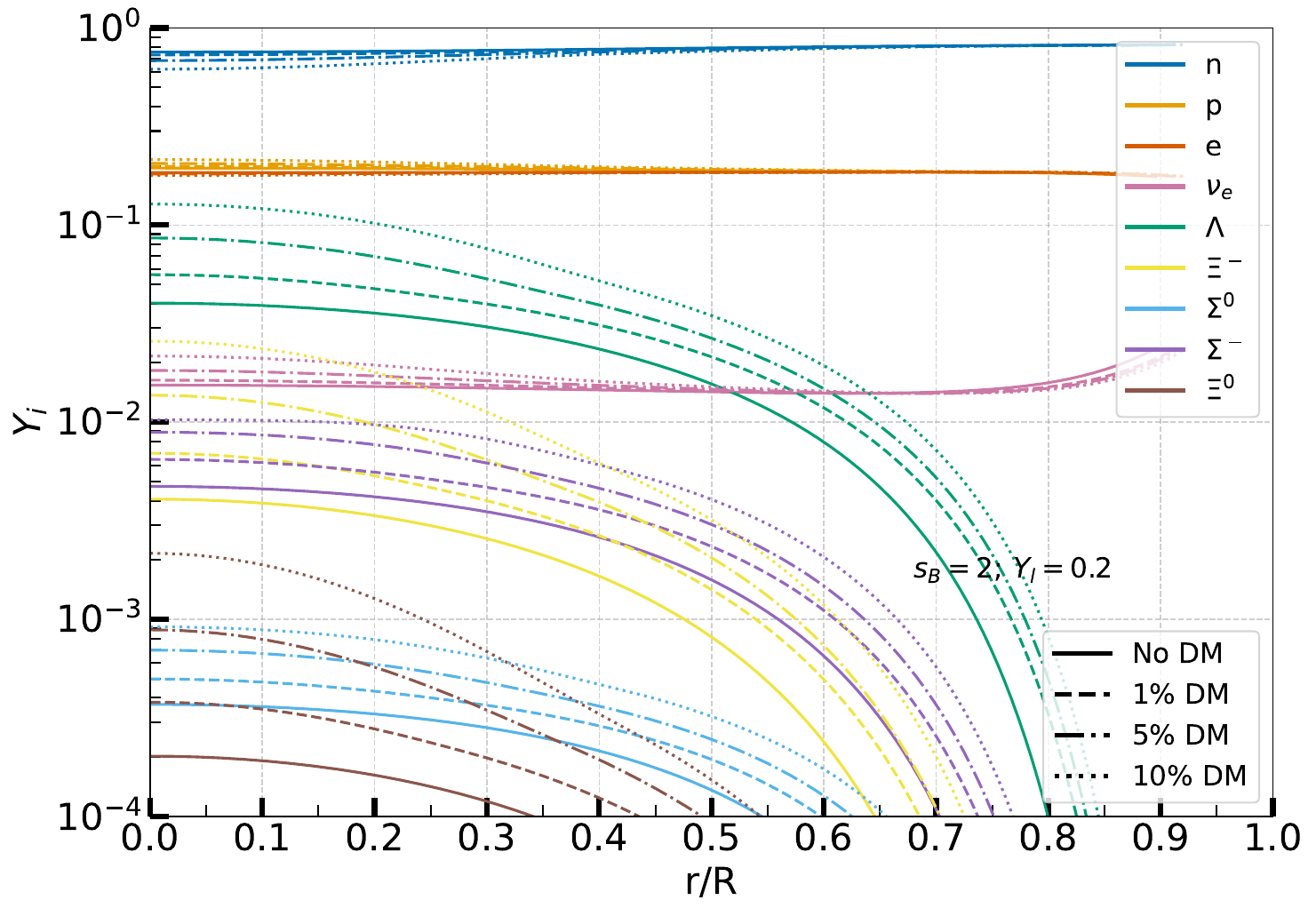}
\includegraphics[width=0.49\linewidth]{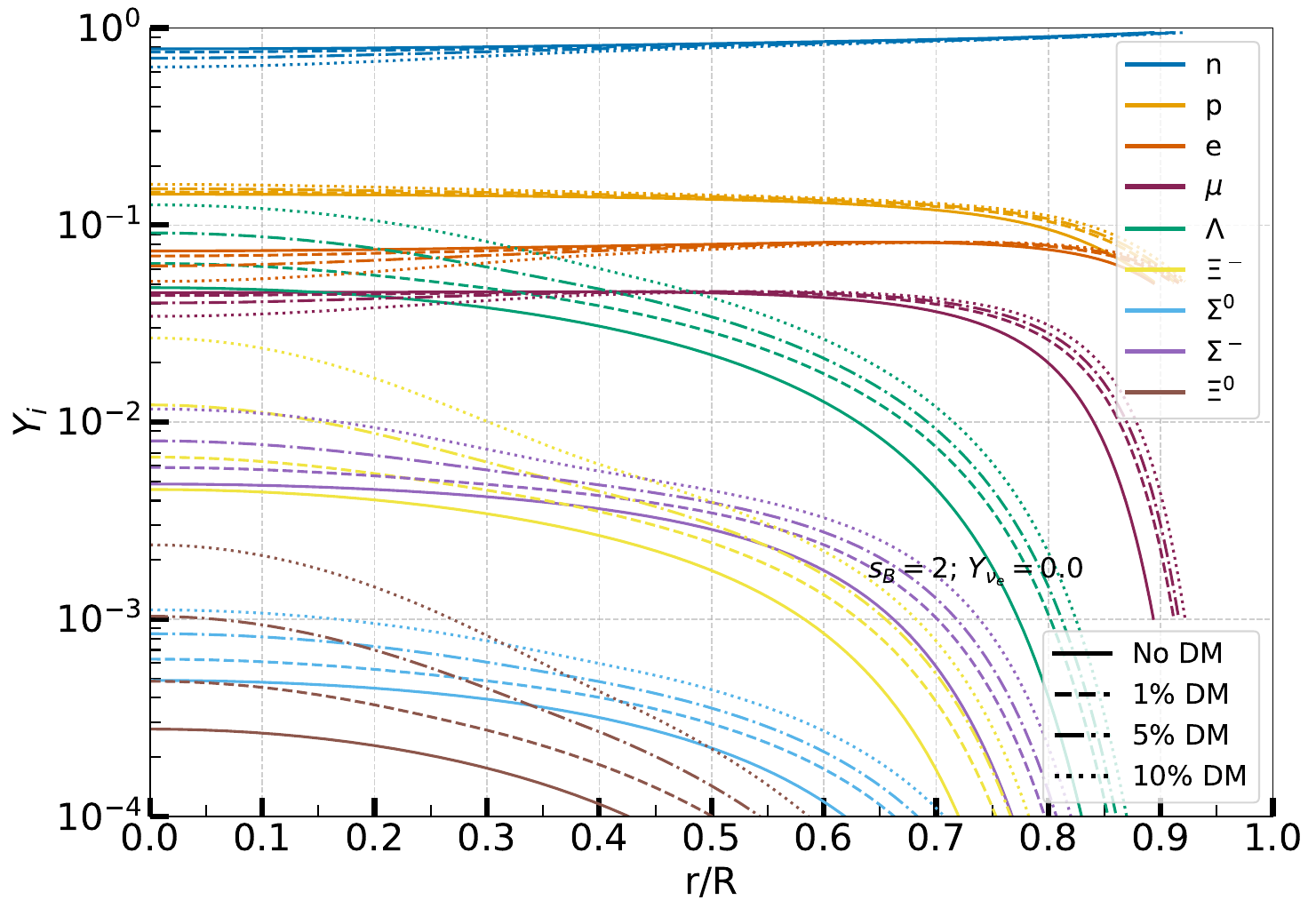} 
\includegraphics[width=0.49\linewidth]{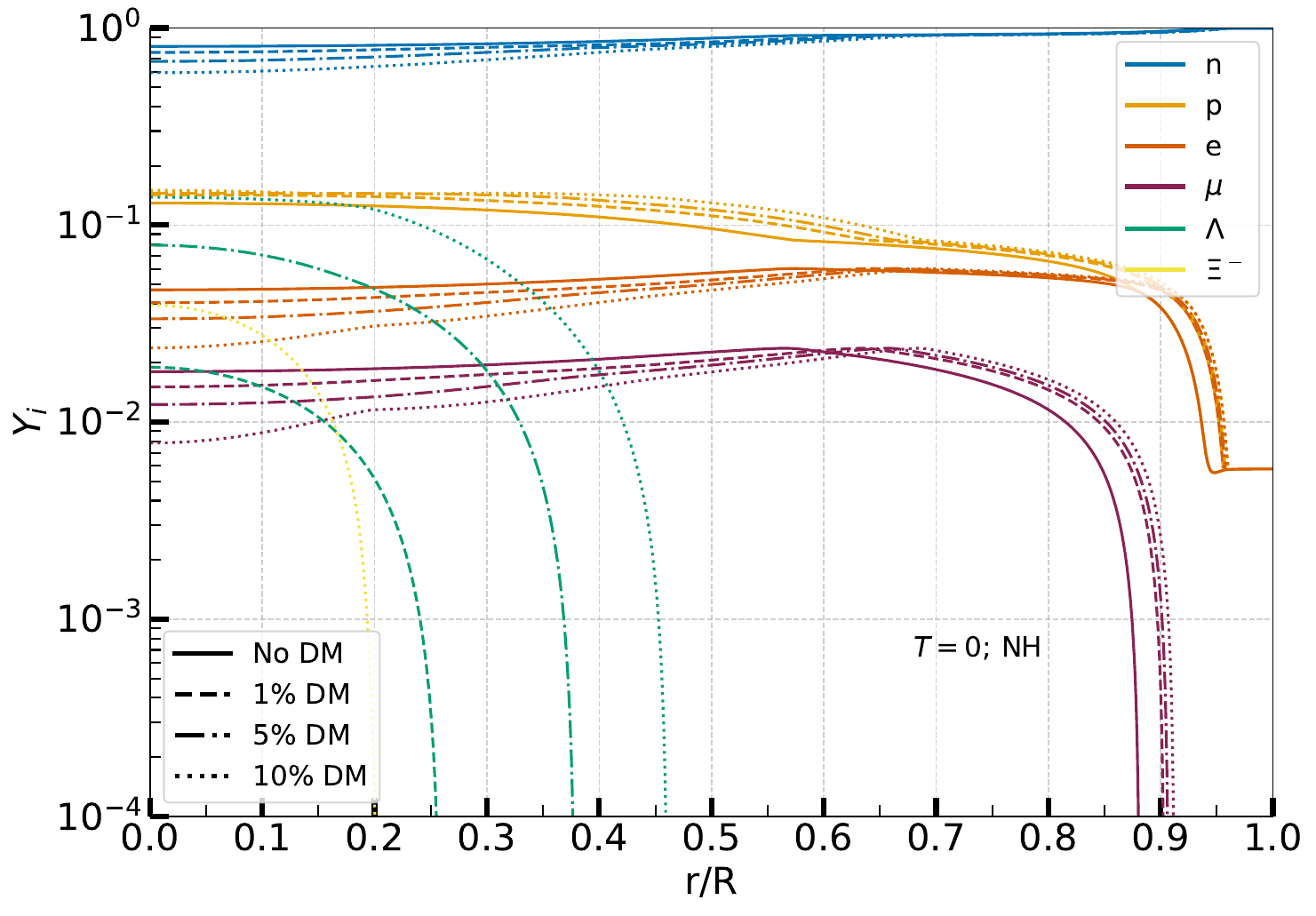}
\caption{Particle distribution profiles versus normalized radius for a PNS with $M_B = 1.55\,M_\odot$, comparing a purely hyperonic EOS with three bosonic DM mass fractions for core configuration (1\%, 5\%, and 10\%). Top panels: neutrino-trapped regime—left: first stage ($s_B = 1,\, Y_l = 0.4$), right: second stage ($s_B = 2,, Y_l = 0.2$). Bottom panels: neutrino-transparent regime; left: third stage ($s_B = 2,\, Y_{\nu_e} = 0$), right: final cold, catalyzed NS ($T = 0$ MeV). 
}
\label{fig:yxr}
\end{figure*}

Figure~\ref{fig:yxr} shows particle distribution profiles as a function of normalized radius and reveals how DM enhances hyperon production across the evolution of a $M_B = 1.55\,M_\odot$ PNS from birth to the cold, catalyzed NS stage, across four snapshots. Four configurations are considered: purely hyperonic EOS (H) without DM (solid lines), and H plus 1\%, 5\%, or 10\% bosonic DM core formation (dashed, dash-dotted, and dotted lines, respectively). In the first snapshot (top-left), DM enhances $\Lambda$ hyperon abundances by softening the EOS, so that a higher central pressure and energy density are needed to sustain the same stellar mass. Nucleons and leptons show little variation due to high neutrino pressure associated with the large lepton fraction, which stabilizes the star. The combination of high $Y_l$ and low $s_B$ keeps the stellar temperature modest (as shown in \cref{fig:temp_radius_core}), suppressing hyperon production relative to later stages.

During deleptonization (top-right), all hyperons appear throughout the stellar matter, with abundances increasing with DM fraction. Higher $s_B$, lower $Y_l$, and heating from the passage of external shock waves after the supernova explosion raise the temperature, favoring thermal production of hyperons. Neutrino fractions increase slightly, while neutron fraction decreases toward the core, indicating that DM promotes neutrino retention, which can affect cooling rates and stellar age since the cooling process is dominated by neutrino emission. The proton fraction also increases with DM growth to balance the increased abundance of negatively charged hyperons in the stellar matter.

In the neutrino-transparent stage (bottom-left), hyperons appear closer to the crust, with core abundances rising with DM content. Muons emerge due to the absence of neutrinos, while electrons and muons decrease slightly towards the core with growing DM fraction. The proton fraction increases with growing DM to maintain charge neutrality due to the growth of negatively charged hyperons. The stellar temperature peaks, promoting widespread hyperon production before the star begins cooling.

Finally, in the cold NS stage (bottom-right), only $\Lambda$ hyperons are abundant for the no DM, 1\%, and 5\% DM cases, while 10\% DM also produces significant $\Xi^-$. With fewer particle species, the impact of DM on the particle abundances is more pronounced at this stage. Leptons and neutrons decrease slightly, and protons increase to preserve charge neutrality due to the negatively charged $\Xi^-$. Comparison across stages indicates that temperature promotes hyperon production and appearance at lower densities, whereas DM amplifies their overall abundance.

\subsection{Mass-Radius Profiles}\label{mr}

\begin{figure*}
    \centering
    \includegraphics[width=0.64\linewidth]{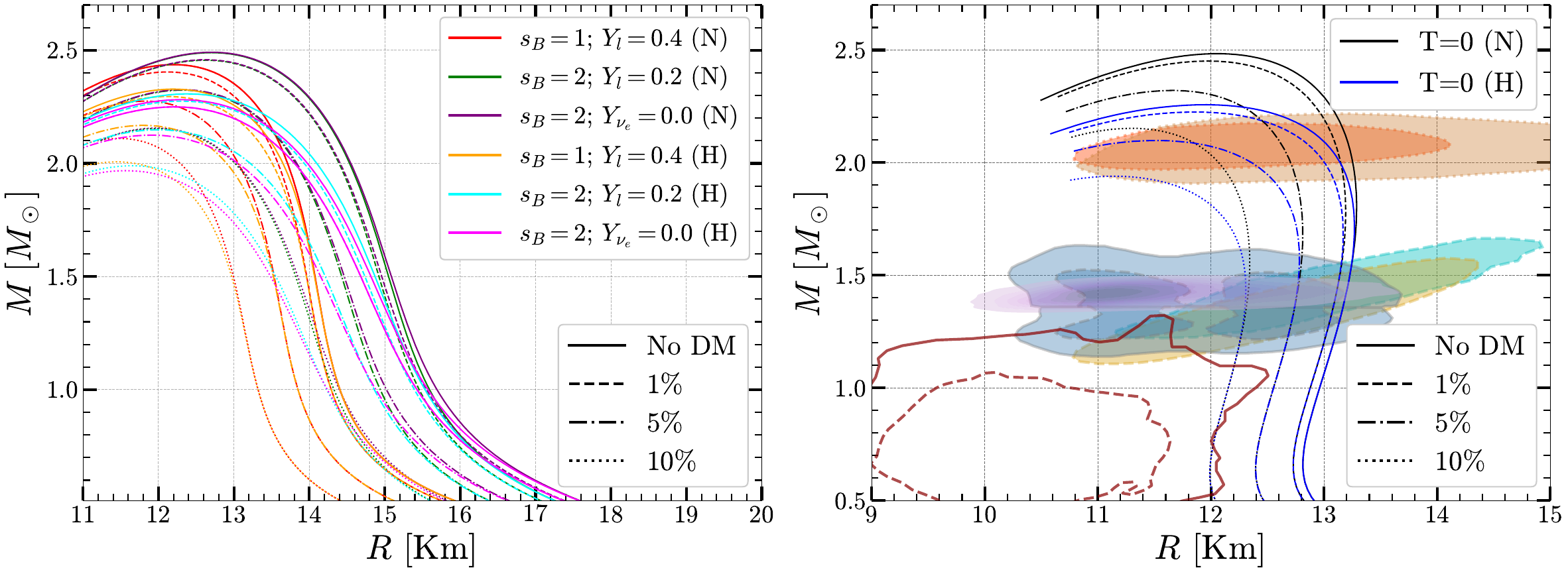}
    \includegraphics[width=0.32\linewidth]{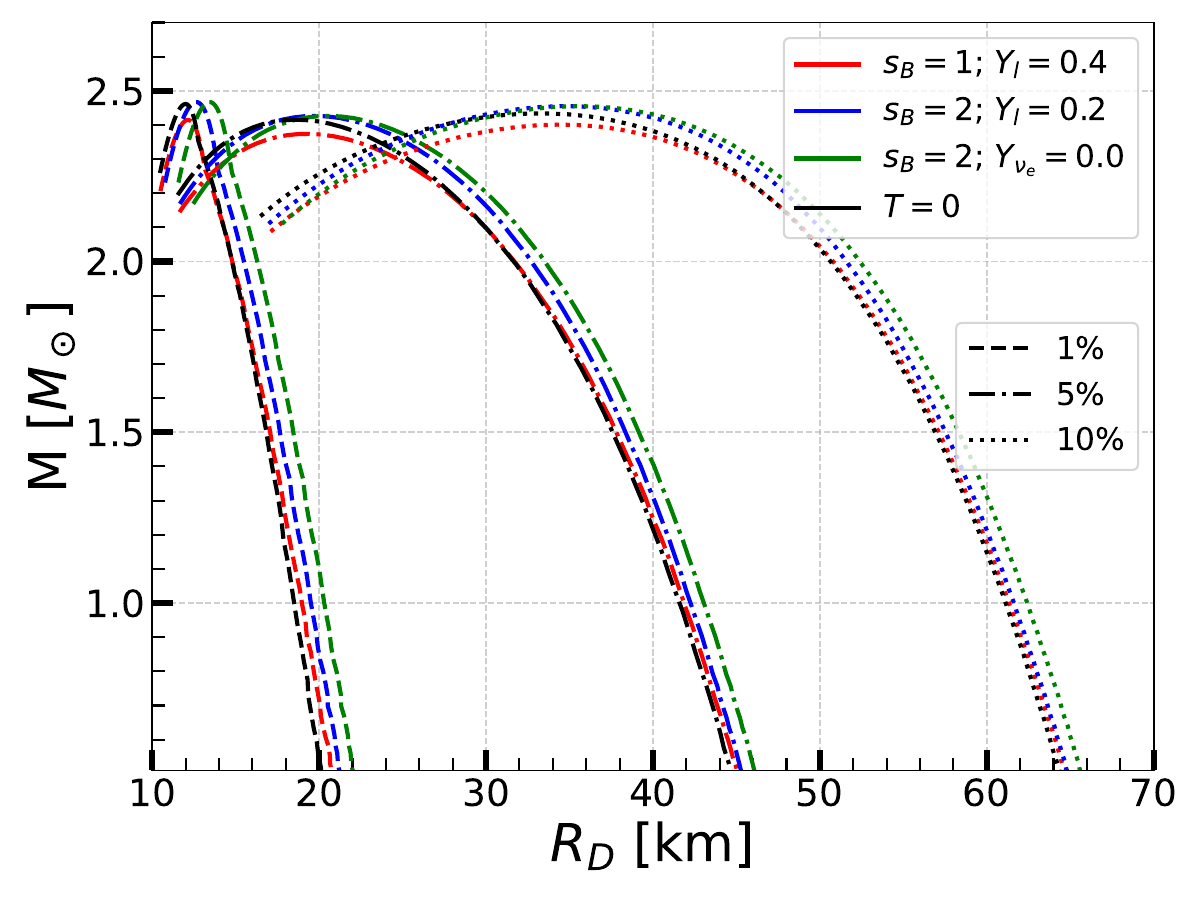}
    \caption{Mass–radius relations for PNS models with and without bosonic DM. The first panel shows hot PNSs configurations, including neutrino-trapped cases with varying $s_B$ and $Y_l$, compared with a neutrino-transparent case ($s_B=2$, $Y_{\nu_e}=0$). Curves are color-coded by composition and styled by DM fraction: no-DM (solid), 1\% (dashed), 5\% (dash-dotted), and 10\% (dotted). The middle panel presents cold ($T=0$ MeV) NS configurations, comparing nucleonic (black) and hyperonic (blue) stars, with line styles indicating $F_D$. Observational constraints include GW170817 (steel blue, 90\% and 50\% CIs) \cite{LIGOScientific:2018hze}, NICER posteriors for PSR J0030+0451 (cyan/yellow) \cite{Riley:2019yda, Miller:2019cac}, PSR J0740+6620 (violet) \cite{Riley:2021pdl, Miller:2021qha}, and PSR J0437–4715 (lilac) \cite{Choudhury:2024xbk}, as well as HESS J1731–347 (red, 90\% and 50\% CL) \cite{2022NatAs...6.1444D}. The last panel shows the DM-halo configuration with $F_D=1\%, 5\%$, and 10\%, where colors denote thermodynamic stages: $s_B=1,\ Y_l=0.4$ (red), $s_B=2,\ Y_l=0.2$ (blue), $s_B=2,\ Y_{\nu_e}=0$ (green), and $T=0$ MeV (black).
  }
    \label{fig:mass_radius_core_bosonic}
\end{figure*}

The right panel of \cref{fig:mass_radius_core_bosonic} shows the mass–radius relations of PNSs at different evolutionary stages, each defined by the $s_B$ and $Y_l$. Results are presented for nucleonic and hyperonic matter, with different color codes distinguishing the compositions according to their thermodynamic conditions, and line styles indicating the fractions of bosonic DM admixed with the star. The solid curves, for example, denote the baseline case without DM. In what follows, we examine the stellar structural changes based on the effects of bosonic DM admixed with PNSs, tracing their evolution through to the cold, catalyzed stage. The first two panels correspond to configurations with DM concentrated in the core, whereas the last panel represents cases where DM forms an extended halo.

At the earliest stage ($s_B = 1$, $Y_l = 0.4$), a hot, neutrino-trapped PNS reaches maximum masses of $\sim 2.44\, M_\odot$ (nucleonic) and $\sim 2.33\, M_\odot$ (hyperonic) without DM. A 5\% DM fraction reduces these to $\sim 2.28\,M_\odot$ and $\sim 2.17\,M_\odot$, respectively, with smaller radii. At later stages ($s_B = 2$, $Y_l = 0.2$ and $s_B = 2$, $Y_{\nu_e} = 0.0$), thermal pressure initially increases both mass and radius, but DM consistently counteracts this by compactifying the star. Even at the hottest, fully deleptonized stage, a 10\% DM fraction lowers the maximum mass to $\sim 2.15\, M_\odot$ for nucleonic stars and below $2.0\, M_\odot$ for hyperonic stars.

Across all evolutionary stages, the presence of DM core systematically shifts the mass-radius relations toward smaller masses and radii. While thermal effects provide outward pressure, the gravitational influence of DM dominates, leading to universally more compact configurations. This effect is amplified in hyperonic stars, as their softer EOS makes them more susceptible to gravitational compression. This increased compactness also promotes the earlier onset of hyperons in the stellar matter (see e.g., \cref{fig:yxr}), further modifying the stellar composition and reinforcing the overall softening of the EOS.

The middle panel of \cref{fig:mass_radius_core_bosonic} shows the mass–radius relations for cold NSs with and without DM. In the absence of DM, nucleonic stars, supported by neutron degeneracy pressure and nuclear interactions, reach a maximum mass of $\sim 2.48\, M_\odot$ with a radius of $\sim 12$ km, while a canonical $1.4\,M_\odot$ star has a radius of $\sim 13.2$ km, reflecting a stiff nucleonic EOS. Including hyperons softens the EOS, reducing the maximum mass to $\sim 2.26\, M_\odot$ and the radius to $\sim 11.9$ km (see \cref{tab:Mmax_R14_H}).

Introducing DM (dashed: 1\%, dash-dotted: 5\%, dotted: 10\%) shifts the $M$–$R$ relations downward and leftward. Unlike baryonic matter, bosonic DM provides no degeneracy pressure and forms a self-gravitating core, increasing the central gravitational pull. This leads to higher central densities and more compact configurations. For instance, in nucleonic stars, 5\% DM lowers the maximum mass to $\sim 2.32\, M_\odot$ with a radius of $\sim 11.7$ km, while 10\% DM reduces these values to $\sim 2.15\, M_\odot$ and $\sim 11.3$ km. Hyperonic stars are more sensitive to DM, with maximum masses dropping below $2.1\, M_\odot$ at higher DM fractions due to their softer EOS.

The last panel shows the effect of the DM halo on the structure evolution of the PNSs. The presence of an extended DM halo fundamentally alters the structural evolution of the PNS. In contrast to the compactifying effect of a central DM core, the halo provides significant external gravitational support, resulting in a star with a substantially larger radius while only marginally increasing the total maximum mass. This occurs because the halo's mass contributes to the overall gravitational field without supplying degeneracy pressure, allowing the baryonic component to expand into a less dense configuration. The effect becomes pronounced during deleptonization, peaking in the neutrino-transparent phase. As neutrino pressure diminishes, the star becomes increasingly sensitive to the halo's gravitational influence, maximizing the resultant expansion once neutrino support is fully lost. Once the stars become cold catalyzed NSs, the halo decreases to a size smaller than when the stars were hot, due to cooling and compactification.

\subsection{Temperature Profiles}\label{tp}

\begin{figure*}[ht!]
		\begin{minipage}[t]{1\textwidth}		 		
  \includegraphics[width=\textwidth]{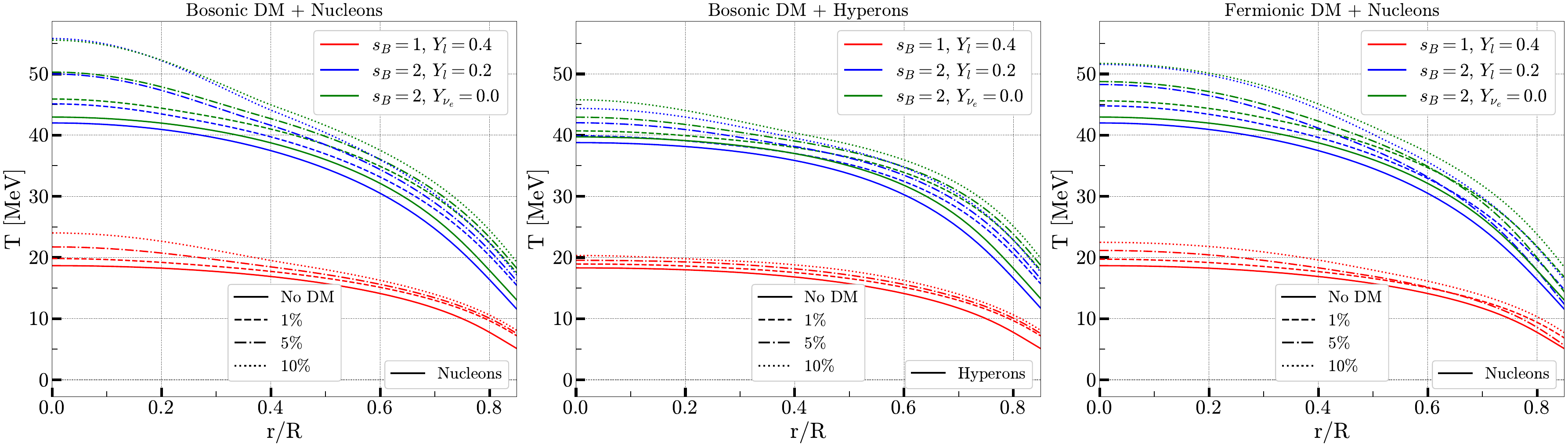}
			 	\end{minipage}
		 		
			 			\caption{Temperature profiles as a function of the normalized radius $r/R$ for PNSs with bosonic and fermionic DM cores. The first panel shows nucleonic stars with a bosonic DM core ($m_\phi = 500$~MeV), the middle panel shows hyperonic stars with a bosonic DM core ($m_\phi = 500$~MeV), and the last panel shows nucleonic stars with a fermionic DM core ($m_D =  939$~MeV). Colors represent thermal and lepton fraction conditions: $s_B = 1,\ Y_l = 0.4$ (red), $s_B = 2,\ Y_l = 0.2$ (blue), and $s_B = 2,\ Y_{\nu_e} = 0.0$ (green). Line styles indicate DM fractions: 1\% (dash-dotted), 5\% (dashed), and 10\% (dotted).
                        }
		\label{fig:temp_radius_core}	 	
     \end{figure*}

{\Cref{fig:temp_radius_core} shows the temperature distribution in PNSs admixed with a fixed fraction of bosonic or fermionic DM core. The figure traces the thermal evolution of a star with baryon mass $M_B = 1.55~\rm M_\odot$ from birth to maturity, enabling us to isolate the thermal response of stellar matter under different evolutionary conditions and to assess the impact of DM on the temperature profile. Since the baryon mass is expected to be conserved throughout the evolution, this approach provides a consistent framework for comparing different stages as well as the influence of distinct DM types on the temperature distribution.

The first panel corresponds to nucleonic stars with varying bosonic $F_D$. In general, the temperature rises with increasing $s_B$ and decreasing $Y_l$. This trend holds across all three panels, whether considering bosonic or fermionic DM in different stellar compositions. Immediately after core birth, when the star is lepton-rich ($Y_l = 0.4$) and $s_B = 1$, the temperature remains relatively modest (red curves). Following core bounce, deleptonization reduces the lepton fraction to $Y_l = 0.2$, while shock heating increases the entropy to $s_B = 2$, producing a hotter configuration (blue curves). After roughly 50 seconds, neutrinos diffuse out of the core ($Y_{\nu_e} = 0$), while the entropy remains fixed at $s_B = 2$. At this stage, the stellar matter reaches its maximum temperature before entering the cooling phase (green curves) \cite{Pons:1998mm, Issifu:2023qoo}.

Beyond the heating driven by thermodynamic conditions, \cref{fig:temp_radius_core} shows that increasing $F_D$ further amplifies the stellar temperature. The admixture of DM increases the star's total mass–energy, strengthening gravity and enhancing compactness \cite{Kain:2021hpk, Issifu:2025gsq}. This deepens the gravitational potential well, compresses the core, and raises the pressure required for hydrostatic equilibrium. The resulting compression reduces interparticle spacing and boosts the gravitational potential energy. By the virial theorem, this is balanced by higher kinetic energy, which appears as additional heating of the stellar matter \cite{Shapiro:1983du, 1978vtsa.book.....C}. These findings are consistent with \cite{Issifu:2024htq}, where a fermionic DM model was studied under thermal equilibrium assumptions.

We emphasize that the term ``DM heating'' used throughout this work refers to a comparative equilibrium effect. At a given evolutionary stage, a PNS containing DM reaches a higher equilibrium temperature than its no-DM or fewer DM admixed configurations at the same baryonic mass. This behavior is structural and gravitational in origin: the presence of DM deepens the gravitational potential well, compresses the baryonic matter, and thereby raises the equilibrium temperature. It should therefore not be interpreted as DM producing a larger rate of temperature increase $dT/dt$ \cite{Giangrandi:2024qdb}, but rather as a shift in the thermal equilibrium configuration driven by gravitational compression.

Comparing the first and last panels highlights a key distinction: bosonic DM produces stronger heating than fermionic DM. The reason lies in quantum statistics. Bosons, unconstrained by the Pauli exclusion principle, can condense into the same ground state, forming a compact, centrally concentrated core that deepens the potential well and strongly compresses nucleonic matter. Fermions, by contrast, are subject to the Pauli principle, which generates degeneracy pressure that resists central crowding. As a result, fermionic DM cores are more extended, produce weaker gravitational deepening, and induce less heating. Altogether, these results suggest that the heating mechanism is robust and generic, arising from the interplay of gravity and thermodynamics, regardless of the precise DM particle type or its degree of thermal coupling with baryonic matter.

A phenomenologically significant observation is that the temperature distribution shows only minimal splitting across different $F_D$ values in the neutrino-rich phase (red curves). However, this splitting more than doubles during the deleptonization and neutrino-transparent stages. This trend is consistent with the results in Tabs.~\ref{tab:Mmax_R14_N} and \ref{tab:Mmax_R14_H}, where the core temperatures for different $F_D$ values are closer in the first stage than in the subsequent two. This behavior highlights the stabilizing role of neutrino pressure in the early, lepton-rich phase, where it mitigates the gravitational influence of DM. As neutrinos diffuse out and the star becomes transparent, the reduction in neutrino pressure allows DM-induced gravitational effects to manifest more strongly, thereby enhancing heating and temperature variation.

The middle panel illustrates the impact of hyperons on the stellar matter admixed with bosonic DM. The inclusion of hyperons increases the particle degrees of freedom, which softens the EOS and reduces the pressure support at a given density \cite{1985ApJ...293..470G, Weissenborn:2011kb, Issifu:2023qyi}. Since the total thermal energy budget of the PNS is conserved, distributing it among a larger number of particles lowers the average thermal energy per particle. Consequently, when comparing the first panel (nucleonic stars admixed with bosonic DM) and the middle panel (hyperonic stars admixed with bosonic DM) panels, the magnitude of the temperature distribution is systematically higher in the nucleonic stars than in those containing hyperons \cite{Malfatti:2019tpg, Issifu:2024fuw, Prakash:1996xs}. Nevertheless, the introduction of a finite $F_D$ produces additional heating, consistent with the trends already discussed for nucleonic stars.

\begin{figure}[t!]
   \centering
	\includegraphics[width=0.45\textwidth]{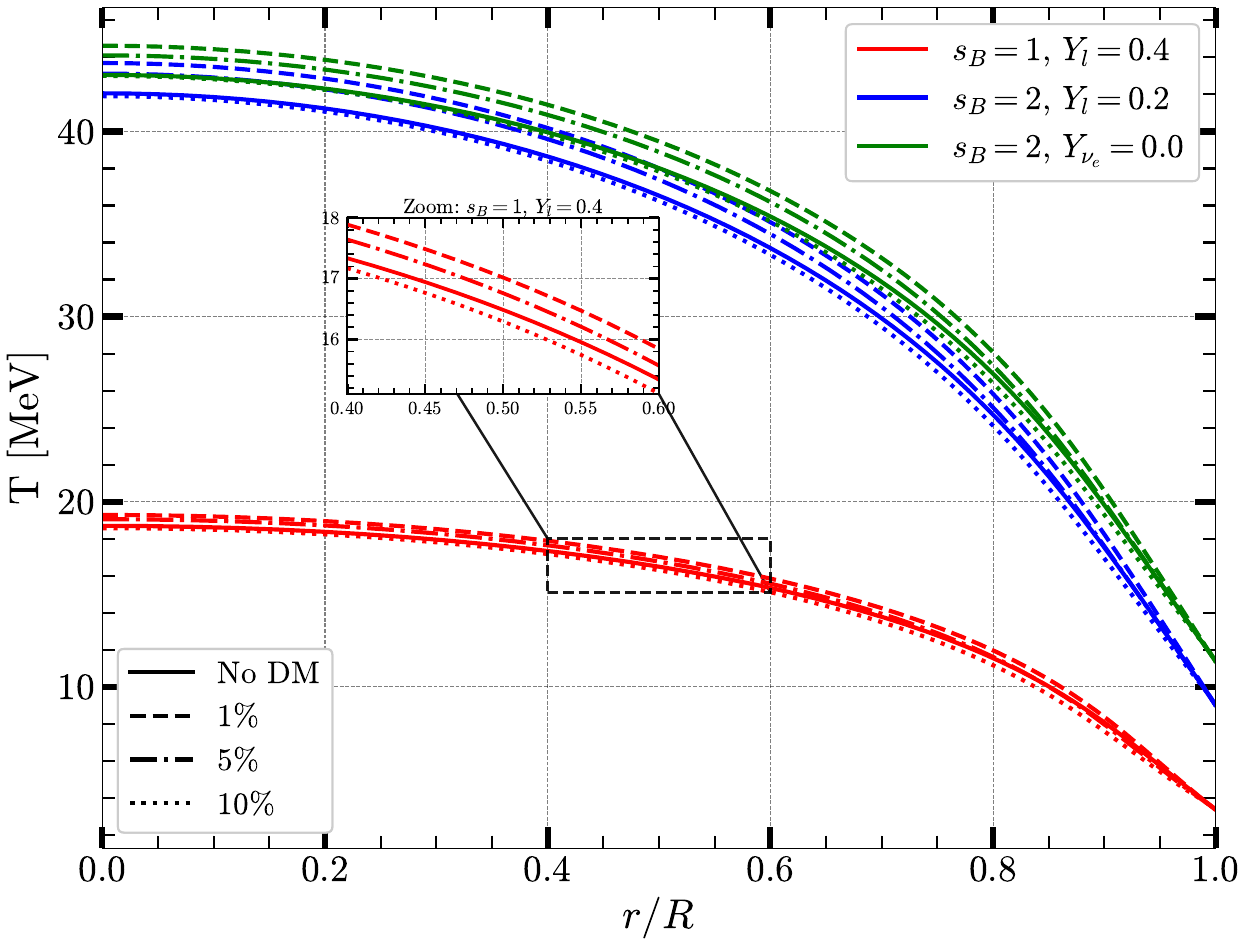}
    	\caption{Radial temperature profiles for DM halo configurations of NSs with varying osonic DM mass fractions (no-DM, 1\%, 5\%, and 10\%). Each color represents a different thermal and lepton configuration: $s_B=1,\,Y_l=0.4$ (red), $s_B=2,\,Y_l=0.2$ (blue), and $s_B=2,\,Y_{\nu_e}=0.0$ (green). The temperature is plotted as a function of normalized radius. The inset zooms into the region $r/R \in [0.4,\,0.6]$ for the case $s_B=1,\ Y_l=0.4$.} 
\label{Temperature_radius_Halo}
	\end{figure}

    \Cref{Temperature_radius_Halo} shows the temperature distribution in an evolving star with a DM halo as a function of the normalized radius. Here, we focus on nucleonic matter, which isolates and captures the general thermodynamic and DM effects on stellar evolution. As discussed above in \cref{fig:temp_radius_core}, the presence of hyperons softens the EOS. It reduces the thermal energy per particle, which in turn lowers the temperature compared to nucleonic matter since the thermal budget of the PNSs is conserved. However, the effects of $s_B$ and $Y_l$ on the stellar matter remain qualitatively similar. Similar to \cref{fig:temp_radius_core} discussed above, the magnitude of the temperature toward the core increases with increasing $s_B$ and decreasing $Y_l$.

    From the third panel of \cref{fig:mass_radius_core_bosonic}, we deduce that as the $F_D$ increases, the star develops a more extended DM halo. As inferred from \cref{Temperature_radius_Halo}, this halo growth correlates with a reduction in the stellar core temperature. The mechanism can be understood as follows: the DM halo deepens the gravitational potential but does not interact thermally with the baryonic sector, which retains all the thermal energy in our framework. Thus, the same baryonic thermal energy is redistributed under stronger gravity, lowering the core temperature. Moreover, because the halo itself provides part of the gravitational support, the baryonic core does not need to raise its central pressure or temperature to maintain hydrostatic balance. In this sense, the halo acts as a cold gravitational reservoir that indirectly cools the star by shifting the burden of support away from the baryonic core. This behavior stands in sharp contrast to the case discussed above, where DM is concentrated in the core (see \cref{fig:temp_radius_core}). In that case, DM directly enhances the central gravitational potential well, increasing the pressure and thermodynamic load on the baryonic matter, which raises the core temperature. The key distinction is therefore the spatial distribution of DM: a diffuse halo cools the star by offloading gravitational support externally, while core-admixed DM heats the star by intensifying the central confinement \cite{Zhou:2025dmy, Issifu:2024htq}.

    The inset in \cref{Temperature_radius_Halo} shows a non-monotonic response of the core temperature to increasing $F_D$ in a DM halo. A small DM fraction ($F_D = 1\%$) raises the core temperature above the no-DM baseline (solid line) because the shallow halo deepens the total gravitational potential, increasing the pressure and thermodynamic load on the baryonic core. As the halo grows ($F_D = 5\%$), its extended mass provides significant external gravitational support, reducing the pressure the baryonic core must sustain for hydrostatic equilibrium; the temperature decreases relative to the 1\% case but remains above the baseline. For $F_D = 10\%$, the halo dominates the gravitational potential, effectively ``shielding'' the core, which now supports only a fraction of the total gravity, and the temperature falls below the no-DM case. This behavior reflects the competition between two effects: potential deepening, which heats the core, and external halo support, which cools it. The core temperature is set by the effective gravitational load on the baryonic matter, progressively reduced as the DM halo grows. This effect is also observed in the second and third stages, as shown in the graph.

\begin{figure*}[htbp!]
		\begin{minipage}[t]{1\textwidth}		 		
  \includegraphics[width=\textwidth]{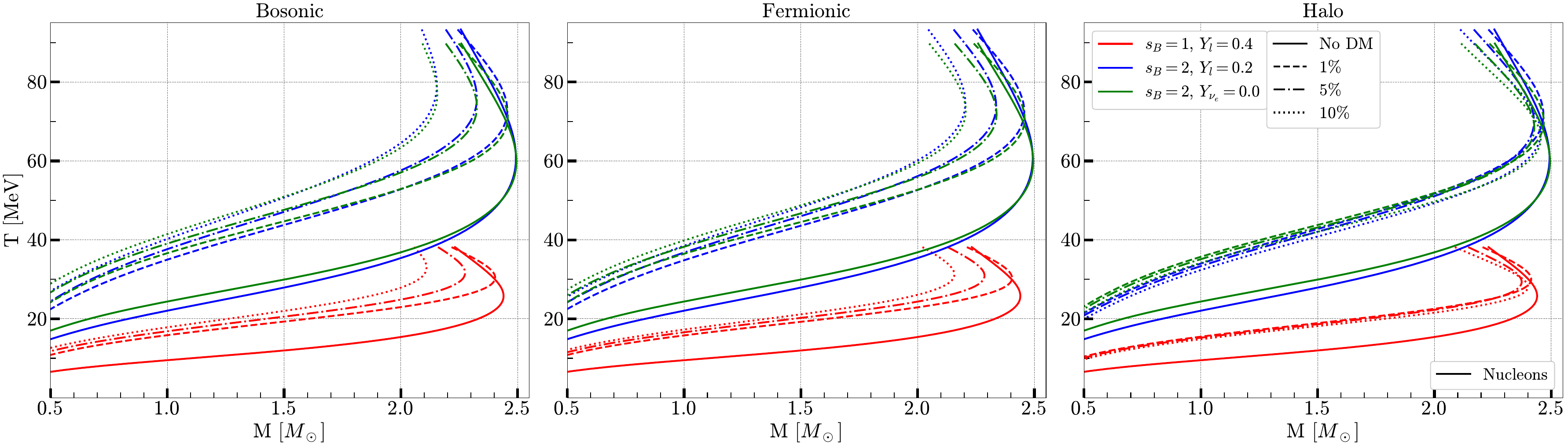}
			 	\end{minipage}
		 		
			 			\caption{Temperature as a function of gravitational mass for different $s_B$ and $Y_l$ conditions. The left panel shows nucleonic stars admixed with a bosonic DM core ($m_\phi = 500$~MeV), the middle panel shows nucleonic stars admixed with a fermionic DM core ($m_D = 939$~MeV), and the right panel shows a nucleonic stars surrounded by an extended bosonic DM halo ($m_\phi = 100$~MeV).
                        }
		\label{fig:temp_mass_fermionic_bosonic}	 	
     \end{figure*}

  \Cref{fig:temp_mass_fermionic_bosonic} shows the central temperature distribution of nucleonic stars as a function of gravitational mass for different DM configurations. Restricting the baryonic sector to nucleons isolates their thermal response to DM, free from complications introduced by hyperons or exotic matter. As discussed earlier (e.g., \cref{fig:temp_radius_core}), hyperons reduce the overall thermal energy but do not alter the qualitative impact of DM fractions. In the core-admixed cases, the temperature rises with stellar mass up to the maximum configuration, reflecting stronger gravitational compression and enhanced DM influence at high central pressures. Beyond the maximum mass, where configurations become unstable, the temperature increases sharply as the mass decreases. 
  
Focusing on the thermal evolution at a fixed mass of $2M_{\odot}$ with a DM fraction of 10\%, which yields the most effective effect in both core and halo configurations, we find a notable trend. Quantitatively, at $F_D=10\%$ in the $s_B = 2,\; Y_l = 0.2$ model (blue curves), the bosonic DM core case reaches 64.55 MeV, exceeding the fermionic (60.14 MeV) and halo (49.31 MeV) cases by approximately 7.3\% and 30.9\%, respectively. In the $s_B = 2,\; Y_{\nu_e} = 0$ configuration (green curves), the fermionic DM core exhibits a temperature of $59.43\,\text{MeV}$, corresponding to a reduction of about $6.7\%$ relative to the bosonic DM core value of $63.39\,\text{MeV}$. In contrast, the bosonic DM halo configuration yields a significantly lower temperature of $49.70\,\text{MeV}$, representing a decrease of approximately $27.6\%$. This pattern reinforces the established hierarchy but reveals a slightly narrower difference between the bosonic and fermionic models compared to the $s_B = 2,\; Y_l = 0.2$ case. Moreover, whereas the bosonic and fermionic DM core configurations show a monotonic increase in temperature with increasing $F_D$, resulting in an overall rise of more than 80\% relative to the no-DM configuration, the halo model displays a clear non-monotonic trend, with the temperature peaking at $F_D = 1\%$ before decreasing at higher fractions. This contrast highlights fundamentally different heating and regulatory mechanisms associated with each class of DM.  

The last panel highlights the contrasting role of an extended DM halo. Unlike core-admixed DM, which consistently heats the baryonic matter by deepening the central potential, a diffuse halo redistributes gravitational support outward. A small halo fraction induces only mild heating, but as the fraction increases, the core temperature decreases and can fall below the no-DM baseline when the halo fraction exceeds $10\%$, as considered here. These results demonstrate that the thermal evolution of nucleonic stars depends sensitively on the DM type, fraction, and spatial distribution, with core DM causing heating and halos producing a relative cooling effect.

The net cooling of the single star with a massive halo (10\% configuration), shown in \cref{Temperature_radius_Halo} for the model with $M_B = 1.55~M_\odot$, arises because the halo's extended mass distribution dominates the gravitational potential. This provides substantial external support, effectively shielding the baryonic core and reducing its pressure and thermodynamic load. The contrasting behavior, net cooling in this single-star case versus the relative cooling trend across stellar sequences, stems from the nature of the comparison. Although the same model is applied in both analyses, the single-star evolution tracks a fixed baryonic mass, whereas the stellar sequences compare different equilibrium configurations. In these sequences, increasing the halo mass progressively shifts gravitational support outward, systematically lowering central compactification and the associated heating as one moves from smaller to larger halo fractions.
     
     }

     \subsection{DM-specific Effects}\label{dms}
 The preceding analysis (Figs.~\ref{fig:Pressure_radius_core_halo}, \ref{fig:mass_radius_core_bosonic}, and Tabs.~\ref{tab:Mmax_R14_N} and \ref{tab:Mmax_R14_H}) demonstrates that core-DM effectively softens the EOS, resulting in increased central baryon density and compactness. Similar effects arise from the inclusion of hyperons (which directly soften the EOS), as both reduce the gravitational mass and shrink the stellar radius \cite{Issifu:2024htq}. This overlap underscores the importance of carefully disentangling DM-induced signatures from those of exotic baryons. Their contrasting impacts on thermal evolution provide a key diagnostic. Hyperonic degrees of freedom reduce the temperature distribution (middle panel of \cref{fig:temp_radius_core}) because the additional degrees of freedom lower the thermal energy per particle \cite{Issifu:2023qyi, Raduta:2020fdn}. In contrast, DM raises the stellar temperature (first and second panels of figs.~\ref{fig:temp_radius_core} and \ref{fig:temp_mass_fermionic_bosonic}) by increasing compactness, deepening the gravitational potential, and enhancing thermal energy through the virial theorem \cite{Carroll_Ostlie_2017}. 
 
The particle fractions in \cref{fig:yxr} indicate that core DM retains neutrinos, whose emission dominates the star’s cooling, while the presence of a DM halo extends the stellar radius and slightly increases the gravitational mass. Hyperons, in contrast, soften the EOS without producing such halo effects, reducing both the stellar radius and gravitational mass \cite{Issifu:2024htq}. Taken together, these results show that although both core DM and hyperons produce similar structural modifications, namely, effective or direct softening of the EOS and increasing central compactness, their thermal signatures are markedly different. Hyperons lower the temperature distribution by introducing additional degrees of freedom that reduce the thermal energy per particle \cite{Issifu:2023qyi, Raduta:2020fdn}, whereas core DM raises the stellar temperature by deepening the gravitational potential and enhancing thermal energy through the virial theorem \cite{Carroll_Ostlie_2017}. Importantly, the heating induced by core DM is a unique effect, absent in hyperonic stars. This provides a clear observational prediction: measurements of neutrino emission from supernova or the early cooling stages of neutron stars could discriminate between the presence of a DM core and hyperonic matter.

The cooling effects associated with hyperon formation and with the presence of a DM halo involve distinct physical mechanisms. When hyperons appear in dense matter, they introduce additional degrees of freedom, which soften the EOS. This typically reduces the maximum mass and radius of the NS and lowers the thermal energy per baryon, which in turn accelerates cooling. In contrast, an extended bosonic DM halo does not soften the baryonic EOS directly; instead, it changes the structure (mass-radius) of the admixed star, which mimics an effective stiffening of the global EOS. In fact, it redistributes gravitational support outward, reducing the core pressure and thermal energy per baryon. This results in a more extended, less tightly bound configuration that enhances neutrino escape and thus promotes cooling. 

Physically, for a fixed $M_B$, an extended DM halo lowers the $n_c$ (see \cref{tab:nucleons_filled}), increasing the neutrino mean free path, $\lambda_\nu \sim (\tilde{\sigma} n_B)^{-1}$, where $\tilde{\sigma}$ is the neutrino scattering cross section, and reducing the diffusion timescale (see \cite{Roberts:2016rsf} for a detailed discussion). Less compact configurations thus cool faster due to the reduced neutrino opacity, consistent with \cite{Pons:1998mm, Prakash:1996xs, Shapiro:1983du}. This effect becomes more pronounced for larger halo fractions, as seen in \cref{Temperature_radius_Halo}, where the $F_D = 10\%$ halo configuration produces net cooling relative to the no-DM baseline for a star with $M_B = 1.55\,M_\odot$. Therefore, comparing the cooling signatures arising from hyperonic matter and DM halo configurations could provide observational clues about the presence and role of DM in compact stars. Such distinctions may be relevant when interpreting anomalously rapid cooling in certain pulsars.

\section{Final Remarks and Conclusion}\label{fr}

A key finding is the identification of a robust DM-induced temperature enhancement in PNSs when DM is concentrated in the core. This gravitational heating arises not from particle annihilation \cite{Bell:2018pkk, McKeen:2023ztq} or an increases in the rate of temperature change ($dT/dt$), but from gravitational compression: DM deepens the stellar core’s gravitational potential, requiring higher baryonic pressure to maintain hydrostatic equilibrium, which, via the virial theorem, raises the equilibrium temperature relative to a DM-free star. Distinct from annihilation-induced heating or cooling from exotic baryons, this effect is systematically explored here for both fermionic and bosonic asymmetric DM, including the first study of BEC DM in PNSs. Using a two-fluid formalism where DM interacts with OM solely through gravity, we trace its imprint throughout the star's evolution, from its hot, neutrino-rich birth to its cold, catalyzed state.

\begin{itemize}
    \item Core-Concentrated DM (Fermionic or Bosonic): As discussed above, this effect is particularly pronounced for bosonic DM, which can form denser, more centrally concentrated cores unconstrained by the Pauli exclusion principle, thereby inducing stronger gravitational compression and larger temperature enhancements compared to its fermionic counterpart.
    
   \item DM Halo Configuration: Conversely, an extended DM halo provides external gravitational support, partially shielding the baryonic core. This results in an initial increase in the core temperature. However, for larger halo fractions ($F_D\,\geq\,5\%$), the additional gravitational support reduces the need for thermal pressure in the core, leading to a relative cooling. This relative cooling effect becomes dominant for larger halo fractions, and can fully reverse the heating induced by the presence of DM as observed when we tracked an individual star of $M_B = 1.55\, M_\odot$ (see \cref{Temperature_radius_Halo}).
\end{itemize}
This thermal signature, core and small halo DM heats, larger halo DM cools, offers a potential observational diagnostic, though similar variations could also arise from EOS uncertainties. Consequently, the diagnostic value lies not in temperature measurement alone, but in the combination of observables: increased compactness, shifted hyperon onset, and enhanced neutrino retention, a pattern unlikely from EOS effects alone. Definitive discrimination requires multi-messenger constraints from supernova neutrino light curves and young pulsar cooling tracks. Unlike hyperons or exotic baryons, which soften the EOS and lower the core temperature by distributing thermal energy over more degrees of freedom, DM uniquely raises the stellar temperature through gravitational compression. Beyond increasing compactness and reducing radius, DM also reshapes temperature and particle distribution profiles, accelerating hyperon onset and amplifying their abundances. During the neutrino-trapped and deleptonization stages, a higher core-DM fraction further enhances neutrino retention, delaying their escape and modifying the early cooling pathway. The combined interplay of DM-induced heating, hyperonic cooling, and altered neutrino transport creates distinctive, stage-dependent imprints on PNS evolution, yielding observationally testable signatures in supernova neutrino light curves and the cooling tracks of young pulsars. 

Moreover, our analysis of radial pressure profiles reveals an interesting trend: core-DM configurations increase central pressure through enhanced compression. In contrast, halo-DM configurations redistribute support outward, lowering the core pressure. These contrasting local pressure effects directly correlate with DM-induced heating versus cooling, as well as with the timing of hyperon onset. Higher core pressure favors DM-induced heating and the early onset of hyperons with an enhanced abundance.

Consequently, deviations in the early cooling curves of NSs, most notably anomalously high temperatures during deleptonization, may indicate the presence of a central DM core. This gravitational heating mechanism provides a distinct, non-annihilating signature of DM, offering a new avenue to probe its presence and distribution through multi-messenger observations of supernova and young pulsars. In the case of a DM halo, a net heating effect is still observed relative to the no-DM baseline due to the star's gravitational response, as shown in \cref{tab:nucleons_filled} and \cref{fig:temp_mass_fermionic_bosonic}. However, larger halo fractions result in comparatively cooler configurations, leading to a relative reduction in temperature with increasing halo mass. Additionally, a net cooling (see \cref{Temperature_radius_Halo}) was observed, for 10\% DM halo, when tracking the evolution of a single star with $M_B = 1.55\, \rm M_\odot$ corresponding to $\sim 1.4\, \rm M_\odot$ gravitational mass. 

The DM-induced heating identified in this work stands in clear contrast to the cooling effect reported in \cite{Issifu:2025qqw}, where DM is thermally coupled to a standard model particle, highlighting the critical role of the DM–OM coupling mechanism. In the single-fluid model of \cite{Issifu:2025qqw}, thermally coupled DM acts as a heat reservoir that cools the stellar system. By contrast, our two-fluid framework, limited to purely gravitational interactions, shows that DM instead heats the star through gravitational compression. This stark dichotomy, heating versus cooling, provides a suitable observational diagnostic: measurements of NS thermal evolution can now potentially discriminate not only the presence of DM but also its fundamental coupling mechanism to OM.

\section*{Acknowledgement}
A.I. acknowledges financial support from the São Paulo State Research Foundation (FAPESP), Grant Nos. 2023/09545-1 and 2025/17347-0. 
This work is part of the project INCT-FNA (Proc. No. 464898/2014-5) and is also supported by the National Council for Scientific and Technological Development (CNPq) under Grants Nos. 303490/2021-7 (D.P.M.) and 306834/2022-7 (T.F.). T. F. also thanks the financial support from  Improvement of Higher Education Personnel CAPES (Finance Code 001) and FAPESP Thematic Grants (2023/13749-1 and 2024/17816-8).  P.~Thakur and Y.~Lim is supported by the National Research Foundation of Korea (NRF) grant funded by the Korea government (MSIT) (No.~RS-2024-00457037). This work was supported (in part) by the Yonsei University Research Fund(Yonsei University Frontier Fellowship for Postdoctoral Researchers) of 2025. Y.~Lim is also supported by Global - Learning \& Academic research institution for Master's and PhD students. D.R.K was in part, supported by Polish NCN Grant No.
2023/51/B/ST9/02798 and SONATINA 7 grant NO. 2023/48/C/ST2/00297. F.M.S. would like to thank CNPq for financial support under research project No. 403007/2024-0 and research fellowship No. 201145/2025-1. The authors acknowledge support by the High Performance and Cloud Computing Group at the Zentrum für Datenverarbeitung of the University of Tübingen, the state of Baden-Württemberg through bwHPC and the German Research Foundation (DFG) for funding under ``Project number 455787709'' (bwForCluster BinAC 2).

\bibliographystyle{apsrev4-112}  
\bibliography{references} 
\end{document}